\documentclass[amsmath,amssymb]{revtex4}
\usepackage{graphicx}
\usepackage{amscd,amsmath,amsthm,amsfonts,amssymb}

\def\PT{$\cal{PT}$}

\def\[{\begin{equation}}
\def\]{\end{equation}}

\advance\textheight -16mm

\numberwithin{equation}{section}

\begin{document}
\title{General rogue waves in the nonlocal \PT-symmetric nonlinear Schr\"{o}dinger equation}
\author{Bo Yang$^{1,2}$ and Jianke Yang$^{2*}$}
\address{
{\small\it $^1$Shanghai Key Laboratory of Trustworthy Computing, East China Normal University, Shanghai 200062, China} \\
{\small\it $^2$Department of Mathematics and Statistics, University of Vermont, Burlington, VT 05401, U.S.A} \\
{\normalsize \small \it * Corresponding author, email address: jyang@math.uvm.edu}}

\begin{abstract}
Rogue waves in the nonlocal \PT-symmetric nonlinear Schr\"odinger (NLS) equation are studied by Darboux transformation. Three types of rogue waves are derived, and their explicit expressions in terms of Schur polynomials are presented. These rogue waves show a much wider variety than those in the local NLS equation. For instance, the polynomial degrees of their denominators can be not only $n(n+1)$, but also $n(n-1)+1$ and $n^2$, where $n$ is an arbitrary positive integer. Dynamics of these rogue waves is also examined. It is shown that these rogue waves can be bounded for all space and time, or develop collapsing singularities, depending on their types as well as values of their free parameters. In addition, the solution dynamics exhibits rich patterns, most of which have no counterparts in the local NLS equation.

\end{abstract}

\maketitle

\section{Introduction}

Rogue waves are spontaneous large waves that ``appear from nowhere and disappear with no trace" \cite{Akhmediev2009}. These waves
attracted a lot of attention in recent years due to their dramatic and often damaging effects, such as in the ocean and optical fibers \cite{KPS2009,SRKJ2007}. The first analytical expression of a rogue wave was derived for the nonlinear Schr\"odinger (NLS) equation by Peregrine in 1983 \cite{PDH1983}. Later, higher-order rogue waves in the NLS equation were derived, and their interesting dynamical patterns were revealed \cite{AAS2009,DGKM2010,ACA2010,DPMB2011,KAAN2011,GLML2012,OhtaJY2012,DPMVB2013}. Since then, analytical rogue waves have been derived for a large number of other integrable systems, such as the derivative NLS equation \cite{XuHW2011,GLML2013}, the three-wave interaction equation \cite{BCDL2013}, the Davey-Stewartson equations \cite{OhtaJKY2012,OhtaJKY2013}, and many others \cite{AANJM2010,OhtaJKY2014,ASAN2010,TaoHe2012,BDCW2012,PSLM2013,Grimshaw_rogue,MuQin2016,LLMSasa2016,LLMFZ2016}. Experimental observations of rogue waves have also been reported in optical fibers and water tanks \cite{KFFMDGAD2010,CHAN2011,CHOSSPA2012}.

Most of the rogue waves derived earlier were for local integrable equations, i.e., the solution's evolution depends only on the local solution value and its local space and time derivatives. Recently, a nonlocal NLS equation
\[ \label{e:PTNLS}
\textrm{i}u_t(x,t)=u_{xx}(x,t)+2\sigma u^2(x,t)u^*(-x,t),
\]
was proposed and studied \cite{AblowitzMussPRL2013,AblowitzMussNonli2016,WYY2016,HXLM2016,Gerdjikov2017}. Here, $\sigma=\pm 1$ is the sign of nonlinearity (with the plus sign being the focusing case and minus sign the defocusing case), and the asterisk * represents complex conjugation. Notice that here, the solution's evolution at location $x$ depends on not only the local solution at $x$, but also the nonlocal solution at the distant position $-x$. That is, solution states at distant locations $x$ and $-x$ are directly related, reminiscent of quantum entanglement between pairs of particles. Eq. (\ref{e:PTNLS}) was called \PT-symmetric since it is invariant under the combined action of the \PT operator, i.e., the joint transformation $x\to -x$, $t\to -t$ and complex conjugation \cite{Yang_review}. Hence, if $u(x,t)$ is a solution, so is $u^*(-x,-t)$. The application of this \PT-symmetric NLS equation for an unconventional system of magnetics was reported in \cite{PTNLSmagnetics}. Following this nonlocal \PT-symmetric NLS equation, many other nonlocal integrable equations were quickly reported and investigated
\cite{AblowitzMussPRE2014,Yan,Khara2015,Zhu1,Fokas2016,Lou,Lou2,AblowitzMussSAPM,Chow,ZhoudNLS,ZhouDS,HePTDS,HePPTDS,JZN2017,Zhu2,Zhu3,AblowitzSG2017,BYnonlocalDS}.
These nonlocal equations are distinctly different from local equations for their novel space and/or time coupling, which could induce new types of solution dynamics and inspire physical applications in nonconventional settings. A connection between nonlocal and local equations was discovered in \cite{BYJY2017}, where it was shown that many nonlocal equations could be converted to local equations through transformations.

Rogue waves in nonlocal integrable equations is an interesting and largely unexplored subject. While rogue waves in the nonlocal Davey-Stewartson equations have been briefly investigated in \cite{HePTDS,HePPTDS,BYnonlocalDS,BYJY2017}, these waves in the more fundamental nonlocal NLS equation (\ref{e:PTNLS}) have received little attention. Since any $x$-symmetric solution of the local NLS equation would also satisfy the nonlocal NLS equation (\ref{e:PTNLS}), $x$-symmetric rogue waves of the local NLS equation, such as the Peregrine wave \cite{PDH1983}, would also be rogue waves of the nonlocal equation (\ref{e:PTNLS}). But whether the nonlocal NLS equation admits $x$-asymmetric rogue waves is the true open question.

In this article, we study rogue waves in the focusing nonlocal NLS equation (\ref{e:PTNLS}) (with $\sigma=1$). We derive these waves by Darboux transformation, and then obtain their explicit algebraic expressions through Schur polynomials. We find three types of rogue waves, which are $x$-asymmetric in general. The first type of solutions has polynomial degrees in their denominators as $n(n+1)$, where $n$ is an arbitrary positive integer. These polynomial degrees match those in the local NLS equation. However, our second and third types of rogue waves have polynomial degrees in their denominators as $n(n-1)+1$ and $n^2$, which have no counterparts in the local NLS equation. This means that the nonlocal NLS equation admits a much wider variety of rogue waves than its local counterpart. Dynamics of these rogue waves is also examined. We show that these rogue waves can be bounded for all space and time. But more importantly, they can also develop collapsing singularities. In addition, the solution dynamics exhibits rich patterns, most of which have no counterparts in the local NLS equation.

\section{General rogue wave solutions}
We consider rogue waves in the focusing nonlocal NLS equation (\ref{e:PTNLS}) (with $\sigma=1$), which approach the same unit constant background when $x, t\to \pm \infty$. In the local NLS equation, general rogue waves in the form of rational solutions of $x$ and $t$ have been reported in \cite{AAS2009,DGKM2010,ACA2010,DPMB2011,KAAN2011,GLML2012,OhtaJY2012,DPMVB2013}, and the degree of the denominator in the $n$-th order rogue wave is $n(n+1)$ \cite{ACA2010,OhtaJY2012}. For the nonlocal NLS equation (\ref{e:PTNLS}), we will show that, in addition to rogue waves with denominator-degree $n(n+1)$, there are also other types of rogue waves with denominator-degrees $n(n-1)+1$ and $n^2$. Thus, the nonlocal NLS equation admits a wider variety of rogue waves than the local NLS equation.

Our results are summarized in the following theorems.

\vspace{0.1cm}
\textbf{Theorem 1}
\emph{The general n-th order type-I rogue waves in the focusing nonlocal NLS equation (\ref{e:PTNLS}) are given by the formula}
\[ \label{N-Rws1}
u_n^{(1)}(x,t)=e^{-2 i t} \left(1+2 \textmd{i} \frac{\tau^{(1)}_1}{\tau^{(1)}_0}\right) ,
\]
\emph{where}
\[
\tau^{(1)}_0= \det_{1\leq i,j \leq n}\left(m^{(1)}_{i,j}\right),\ \ \tau _1^{(1)}=\det\left(   \begin{array}{cc}
                        \left(m^{(1)}_{i,j}\right)_{1\leq i,j \leq n} & \nu^{(1)} \\
                        \mu^{(1)} & 0
                      \end{array}
\right),
\]
\[ \label{detmij1}
m^{(1)}_{i,j}=\lim_{\widetilde{\epsilon}, \epsilon\rightarrow0}\frac{1}{(2i-2)!(2j-2)!}\frac{\partial^{2i+2j-4}}{\partial \widetilde{\epsilon}^{2i-2}\partial \epsilon^{2j-2}}
\left[\frac{\psi(\zeta)\phi(\lambda)}{\lambda-\zeta}\right],
\]
\[
\mu^{(1)}= \left[\phi^{(0)}_1, \phi^{(1)}_1,..., \phi^{(n-1)}_1\right],\
\phi^{(k)}_1=\lim_{\epsilon\rightarrow0}\frac{\partial^{2k}\phi_1(\lambda)}{(2k)! \partial \epsilon^{2k}}, \nonumber
\]
\[
\nu^{(1)} =  \left[ \psi^{(0)}_2, \\ \psi^{(1)}_2 \\ ,...,\\
\psi^{(n-1)}_2 \right]^T,\
\psi^{(k)}_2=\lim_{\widetilde{\epsilon}\rightarrow0}\frac{\partial^{2k}\psi_2(\zeta)}{(2k)! \partial \widetilde{\epsilon}^{2k}}, \nonumber
\]
\begin{eqnarray}
\lambda= \textmd{i}(1+\epsilon^2),\ \zeta=-\textmd{i}(1+\widetilde{\epsilon}^2),
\end{eqnarray}
\emph{and the superscript `$T$' represents the matrix transpose. Here, functions $\phi(\lambda)=(\phi_1, \phi_2)^T$ and $\psi(\zeta)=(\psi_1, \psi_2)$ are defined as}
\begin{eqnarray} \label{Eigenfunction}
&& \phi(\lambda)= \frac{1}{\sqrt{h-1}} \left(
                    \begin{array}{c}
 \sinh\left[A+ \frac{1}{2}\ln\left(h+\sqrt{h^2-1}\right)\right] \\
 \sinh \left[-A+ \frac{1}{2}\ln\left(h+\sqrt{h^2-1}\right)\right]  \\
                    \end{array}
                  \right), \\ \nonumber
&&\lambda=\textmd{i}h, \ h=1+\epsilon^2, \ A=\sqrt{h^2-1} (x-2 \textrm{i} h t+ \textrm{i}\theta ), \ \theta=\sum_{k=0}^{n-1}s_{k}\epsilon^{2k},
\end{eqnarray}
\begin{eqnarray}\label{AdEigenfunction}
&& \psi(\zeta)= \frac{1}{\sqrt{\widehat{h}-1}} \left[
                    \begin{array}{c}
  \sinh \left[ \widehat{A}+\frac{1}{2}\ln\left(\widehat{h}+\sqrt{\widehat{h}^2-1}\right)\right]\\
  \sinh \left[-\widehat{A}+\frac{1}{2}\ln\left(\widehat{h}+\sqrt{\widehat{h}^2-1}\right) \right] \\
                    \end{array}
                  \right]^T,\\ \nonumber
&&\zeta=-\textmd{i}\widehat{h}, \ \widehat{h}=1+\widetilde{\epsilon}^2,\
\widehat{A}=\sqrt{\widehat{h}^2-1} (x+2 \textrm{i} \widehat{h} t+\textrm{i}\widehat{\theta} ),\     \widehat{\theta}=\sum_{k=0}^{n-1}r_{k}\widetilde{\epsilon}^{2k},
\end{eqnarray}
\emph{and $s_{k}$ and $r_{k}\ (k=0,1,\ldots,n-1)$ are free real parameters.}

\vspace{0.1cm}
\textbf{Theorem 2} \emph{The general n-th order type-II rogue waves in the focusing nonlocal NLS equation (\ref{e:PTNLS}) are given by the formula}
\[ \label{N-Rws2}
u_n^{(2)}(x,t)=e^{-2 \textrm{i}\hspace{0.02cm} t} \left(1+2 \hspace{0.02cm} \textrm{i}  \frac{ \tau^{(2)}_1}{\tau^{(2)}_0}\right) ,
\]
\emph{where}
\[
 \tau^{(2)}_0= \det_{1\leq i,j \leq n}\left( m^{(2)}_{i,j}\right),\ \ \tau^{(2)}_1=\det\left(   \begin{array}{cc}
                        \left(m^{(2)}_{i,j}\right)_{1\leq i,j \leq n} & \nu^{(2)} \\
                        \mu^{(2)} & 0
                      \end{array}
\right),
\]
\[ \label{detmij2}
m^{(2)}_{i,j}=\left\{
                            \begin{array}{ll}
                              \lim_{\epsilon\rightarrow0}\frac{1}{(2j-2)!}\frac{\partial^{2j-2}}{\partial \epsilon^{2j-2}}
\left[\frac{\psi_{0} \phi(\lambda)}{\mathbf{i}+\lambda}\right], & \emph{when}\ i=1,\\
                              m^{(1)}_{i-1,j}, & \emph{when}\ i\neq1;
                            \end{array}
                          \right.
\]
\[
\mu^{(2)}=\mu^{(1)},\quad  \nu^{(2)}=\left[-1, \nu_1^{(1)}, \dots, \nu_{n-1}^{(1)}\right]^T,
\]
$\psi_0=\left(1, -1\right)$, and $m^{(1)}_{i,j}$, $\mu^{(1)}$, $\nu^{(1)}$, $\phi$ and $\lambda$ are as given in Theorem 1.

\vspace{0.1cm}
\textbf{Theorem 3}
\emph{The general n-th order type-III rogue waves in the focusing nonlocal NLS equation (\ref{e:PTNLS}) are given by the formula}
\[ \label{N-Rws3}
u_n^{(3)}(x,t)=e^{-2 \textmd{i} t} \left(1+2 \hspace{0.02cm}\textmd{i} \frac{\tau^{(3)}_1}{\tau^{(3)}_0}\right),
\]
\emph{where}
\[
\tau^{(3)}_0= \det_{1\leq i,j \leq n}\left(m^{(3)}_{i,j}\right),\ \ \tau^{(3)}_1=\det\left(   \begin{array}{cc}
                        \left(m^{(3)}_{i,j}\right)_{1\leq i,j \leq n} & \nu^{(3)} \\
                        \mu^{(3)} & 0
                      \end{array}
\right),
\]
\[ \label{detmij3}
m^{(3)}_{i,j}=\lim_{\widetilde{\epsilon}, \epsilon\rightarrow0}\frac{1}{(2i-2)!(2j-2)!}\frac{\partial^{2i+2j-4}}{\partial \widetilde{\epsilon}^{2i-2}\partial \epsilon^{2j-2}}
\left[\frac{\omega(\zeta)\phi(\lambda)}{\lambda-\zeta}\right],
\]
\[
\mu^{(3)}=\mu^{(1)}, \quad
\nu^{(3)} = \left(\omega^{(0)}_2, \\ \omega^{(1)}_2 \\ ,...,\\
\omega^{(n-1)}_2 \right)^T,\
\omega^{(k)}_2=\lim_{\widetilde{\epsilon}\rightarrow0}\frac{\partial^{2k}\omega_2(\zeta)}{(2k)! \partial \widetilde{\epsilon}^{2k}}, \nonumber
\]
\emph{$\omega(\zeta)=(\omega_1, \omega_2)$ is defined as}
\begin{eqnarray}\label{AdEigenfunction3}
&&  \omega(\zeta)=\left(
                    \begin{array}{c}
\cosh \left[ \widehat{A}+ \frac{1}{2}\ln\left(\widehat{h}+\sqrt{\widehat{h}^2-1}\right) \right]\\
-\cosh \left[\widehat{A}- \frac{1}{2}\ln\left(\widehat{h}+\sqrt{\widehat{h}^2-1}\right)\right] \\
                    \end{array}
                  \right)^T,
\end{eqnarray}
\emph{and $\phi$, $\mu^{(1)}$, $\zeta$, $\lambda$, $\zeta$, $\widehat{A}$, $\widehat{h}$ are as given in Theorem 1.}
\vspace{0.2cm}

Rogue waves are rational solutions, and the $\tau_0$ and $\tau_1$ functions in the above theorems are polynomials of $x$ and $t$. More explicit algebraic expressions for these functions can be provided through Schur polynomials $S_n(\textbf{x})$, which are defined by
\begin{eqnarray}\label{Schurpoly}
&& \sum_{k=0}^{\infty}S_{k}(\textbf{x})\epsilon^{k}=\exp\left(\sum_{k=0}^{\infty} x_{k}\epsilon^{k}\right),
\end{eqnarray}
where  $\textbf{x}=\left( x_{1}, x_{2},\cdots  \right)$. Specifically,
\begin{eqnarray}
\nonumber
 && S_{0}(\textbf{x})=1,\ \ S_{1}(\textbf{x})=x_{1},\ \ S_{2}(\textbf{x})=x_{2}+\frac{x_{1}^2}{2},\\ \nonumber
 &&  S_{n}(\textbf{x}) =\sum_{l_{1}+2l_{2}+\cdots+kl_{k}=n} \left( \ \prod _{j=1}^{k} \frac{x_{j}^{l_{j}}}{l_{j}!}\right).
\end{eqnarray}
Using Schur polynomials, we get the following more explicit expressions for these rogue waves.

\vspace{0.1cm}
\textbf{Theorem 4}
\emph{The matrix elements for rogue waves in Theorems 1-3 have the following explicit algebraic expressions,}
\begin{eqnarray*}
&&  m^{(1)}_{i,j}=\sum_{\nu=0}^{\min\{i-1,j-1\}}  \frac{1}{2 \hspace{0.02cm} \textrm{i}} \left(\widetilde{\mathcal{A}}_{2i-1,\nu}\mathcal{A}_{2j-1,\nu}
+\widetilde{\mathcal{B}}_{2i-1,\nu} \mathcal{B}_{2j-1,\nu}\right), \\
&& m^{(2)}_{1,j}= \frac{\textrm{1}}{2\textrm{i}}  \left[S_{2j-1}(\textbf{\emph{W}}^{+})-S_{2j-1}(\textbf{\emph{W}}^{-})\right], \\
&& m^{(3)}_{i,j}=\sum_{\nu=0}^{\min\{i-1,j-1\}}  \frac{1}{2 \hspace{0.02cm} \textrm{i}}   \left(\widetilde{\mathcal{A}}_{2i-2,\nu}\mathcal{A}_{2j-1,\nu}
-\widetilde{\mathcal{B}}_{2i-2,\nu} \mathcal{B}_{2j-1,\nu}\right),  \\
&& \mu_{j}^{(1)}=   S_{2j-1}(\textbf{\emph{Y}}^{+}),\  \nu_{j}^{(1)}= S_{2j-1}(\widetilde{\textbf{\emph{Y}}}^{-}), \ \nu_{j}^{(3)}=-  S_{2j-2}(\widetilde{\textbf{\emph{Y}}}^{-}),
\end{eqnarray*}
\emph{where}
\begin{eqnarray*}
&&  \widetilde{\mathcal{A}}_{i,\nu}=\frac{1}{2^{\nu}} S_{i-2\nu}(\widetilde{\textbf{\emph{X}}}^{+}), \quad
\mathcal{A}_{j,\nu}=\frac{1}{2^{\nu}} S_{j-2\nu}(\textbf{\emph{X}}^{+}), \\
&&  \widetilde{\mathcal{B}}_{i,\nu}=\frac{1}{2^{\nu}} S_{i-2\nu}(\widetilde{\textbf{\emph{X}}}^{-}),\quad
\mathcal{B}_{j,\nu}=\frac{1}{2^{\nu}} S_{j-2\nu}(\textbf{\emph{X}}^{-}), \\
&& \textbf{\emph{X}}^{\pm}=\left(X_{1}^{\pm}, X_{2}^{\pm},\ldots \right),  \
\widetilde{\textbf{\emph{X}}}^{\pm}=\left(\widetilde{X}_{1}^{\pm}, \widetilde{X}_{2}^{\pm}, \ldots, \right),   \\
&& \textbf{\emph{Y}}^{\pm}=\left(Y_{1}^{\pm}, Y_{2}^{\pm},\ldots \right), \ \widetilde{\textbf{\emph{Y}}}^{\pm}=\left(\widetilde{Y}_{1}^{\pm}, \widetilde{Y}_{2}^{\pm}, \ldots, \right), \ \emph{\textbf{W}}^{\ \pm}=\left(W_{1}^{\pm}, W_{2}^{\pm},\ldots \right), \\
&&X_{2k+1}^{\pm}=\sqrt{2}\left[  \sum_{j=0}^k \pm  (\delta_{j,0} x - 2\textrm{i}t  + \textrm{i} s_{j}){\frac{1}{2} \choose \textit{k-j}}
 \left(\frac{1}{2}\right)^{k-j}+\frac{(2k)!}{2^{3k+1}(k!)^2}\frac{\left(-1\right)^{k}}{(2k+1)}\right], \\
&&\widetilde{X}_{2k+1}^{\pm}=\sqrt{2}\left[\sum_{j=0}^k \pm  ( \delta_{j,0} x + 2\textrm{i}t + \textrm{i}r_{j}){\frac{1}{2} \choose \textit{k-j}}
\left(\frac{1}{2}\right)^{k-j}+\frac{(2k)!}{2^{3k+1}(k!)^2}\frac{\left(-1\right)^{k}}{(2k+1)}\right],\\
&& W_{2k+1}^{\pm}=Y_{2k+1}^{\pm}=X_{2k+1}^{\pm},\quad  \widetilde{Y}_{2k+1}^{\pm}=\widetilde{X}_{2k+1}^{\pm}, \quad
W_{2k}^{\pm} = \frac{(-1)^{k}}{ k \cdot 2^{k}},\\
&&X_{2k}^{\pm}(\nu)=\widetilde{X}_{2k}^{\pm}(\nu)=\left(\nu+1\right)\left[\frac{(-1)^{k}}{ k \cdot 2^k}\right], \quad Y_{2k}^{\pm}=\widetilde{Y}_{2k}^{\pm}=0,
\end{eqnarray*}
\emph{$\delta_{j,0}$ is the standard Kronecker delta notation (i.e., $\delta_{0,0}=1$ and zero otherwise),}
$$
{\alpha \choose n}\equiv \frac{\alpha(\alpha-1)\cdots(\alpha-n+1)}{n!},
$$
\emph{and parameters $s_{k}$ and $r_{k}$ are free real constants.}

\vspace{0.15cm}
\textbf{Theorem 5} \emph{All the three types of rogue waves in the above theorems satisfy the following boundary conditions,}
\[
u_n(x,t) \to (-1)^n e^{-2\textmd{i}t}, \quad x, t \to \pm \infty.
\]
\emph{In addition, the degrees of their denominator polynomials are}
\[
\mbox{deg}(\tau^{(1)}_0)=n(n+1), \quad \mbox{deg}(\tau^{(2)}_0)=n(n-1)+1, \quad \mbox{deg}(\tau^{(3)}_0)=n^2.
\]

\vspace{0.2cm}
\textbf{Remark 1.}  Type-I and type-III rogue waves have $2n$ real parameters, $\{s_{k}, r_{k}, 0\le k\le n-1\}$. Type-II rogue waves have $2n-1$ real parameters, $\{s_{0}, s_{1},\ldots, s_{n-1}\}$ and $\{r_{0}, r_{1},\ldots, r_{n-2}\}$. However, the parameter $r_{n-2}$ (when $n\ge 2$) automatically vanishes from this solution. The reason is that $r_{n-2}$ only appears in the last rows of the determinants $\tau_0^{(2)}$ and $\tau_1^{(2)}$. We can readily show that the derivatives of these last rows with respect to $r_{n-2}$ are proportional to the first rows of those determinants; thus the solution $u_n^{(2)}(x,t)$ is actually independent of $r_{n-2}$. This means that type-II rogue waves have only $2n-2$ real parameters $\{s_{0}, s_{1},\ldots, s_{n-1}\}$ and $\{r_{0}, r_{1},\ldots, r_{n-3}\}$ when $n\ge 2$ (they have one real parameter $s_0$ when $n=1$). Utilizing time-translation invariance of the nonlocal NLS equation (\ref{e:PTNLS}), one of those parameters can be further removed for each type of these solutions.

\section{Derivation of rogue-wave solutions}

In this section, we derive the general rogue-wave solutions given in section 2. This derivation is based on the generalized Darboux transformation (DT) first proposed in \cite{MatveevDT1991} and further developed in \cite{GLML2012,Cieslinski2009}. The outline of our derivation is as follows.

We begin with the following ZS-AKNS scattering problem \cite{Ablowitz1981,Zakharov1984}:
\begin{eqnarray}
 &&\Phi_{x}=U(Q,\lambda) \Phi,  \label{Laxpairxp}\\
 &&\Phi_{t}=V(Q,\lambda) \Phi,  \label{Laxpairtp}
\end{eqnarray}
where,
\begin{eqnarray}
&& U(Q,\lambda)= -\textmd{i} \lambda \sigma_{3} + Q,   \label{Uform} \\
 && V(Q,\lambda)=2\textmd{i} \lambda^2 \sigma_{3} - 2\lambda Q  - \textmd{i} \sigma_{3} \left( Q_{x}-Q^2 \right),   \label{Vform} \\
 \vspace{0.2cm}
&&  Q(x,t)=\left(
\begin{array}{cc}
0 & u(x,t) \\
v(x,t) & 0 \\
\end{array}
\right), \quad \sigma_{3}=\textmd{diag}(1,-1).    \label{Qform}
\end{eqnarray}
The compatibility condition of these equations is the zero-curvature equation
\[ \label{zero_curvature}
U_t-V_x+[U, V]=0,
\]
which yields the following coupled system for potential functions $(u, v)$ in the matrix $Q$:
\begin{eqnarray}
&& \textmd{i}u_t=u_{xx}-2u^2v, \label{uequation} \\
&& \textmd{i}v_t=-v_{xx}+2v^2u. \label{vequation}
\end{eqnarray}
The focusing nonlocal NLS equation (\ref{e:PTNLS}) is obtained from the above coupled system under the symmetry reduction \cite{AblowitzMussPRL2013}:
\[ \label{qS}
v(x,t)= -u^*(-x,t).
\]
Under this reduction, the potential matrix $Q$ satisfies the following symmetry condition:
\[ \label{QS}
\sigma_1Q^*(x,t)\sigma_1=-Q(-x,t).
\]
To construct the Darboux transformation, we also introduce the adjoint spectral problem
\begin{eqnarray}
&& \Psi_{x}=-\Psi U(Q,\lambda),  \label{AdLaxpairxp} \\
&& \Psi_{t}=-\Psi V(Q,\lambda).  \label{AdLaxpairtp}
\end{eqnarray}

\subsection{$N$-fold Darboux transformation and its reduction}

For the ZS-AKNS scattering problem (\ref{Laxpairxp})-(\ref{Laxpairtp}) with general potential functions $u(x,t)$ and $v(x,t)$, its Darboux transformation as given in \cite{GLML2012,HXLM2016,Cieslinski2009} is
\begin{eqnarray}\label{eDT}
T=I+\frac{\zeta_{1}-\lambda_{1}}{\lambda-\zeta_{1}}P_{1}, \quad
 P_{1}=\frac{\Phi_{1}\Psi_{1}}{\Psi_{1}\Phi_{1}},
\end{eqnarray}
where $\Phi_{1}$ is a column-vector solution of the original Lax-pair system (\ref{Laxpairxp})-(\ref{Laxpairtp})
with spectral parameter $\lambda=\lambda_{1}$, and $\Psi_{1}$ is a row-vector solution of the adjoint Lax-pair system
(\ref{AdLaxpairxp})-(\ref{AdLaxpairtp}) with spectral parameter $\lambda=\zeta_{1}$. This Darboux transformation closely mimics the dressing matrix in the Riemann-Hilbert formulation of the inverse scattering transform for the focusing NLS equation \cite{Zakharov1984,Yang2010}.
Under this Darboux transformation, if $\Phi(x, \lambda)$ satisfies the original ZS-AKNS scattering equations (\ref{Laxpairxp})-(\ref{Laxpairtp}), then the new function
\[
\Phi_{[1]}=T\Phi
\]
would satisfy the same ZS-AKNS scattering problem,
except that the potential matrix $Q$ is transformed to
\[ \label{DTP}
Q_{[1]}=Q+\textrm{i}(\zeta_{1}-\lambda_{1})\left[  \sigma_{3}, P_{1}  \right].
\]
This relation between the old and new potentials is the B\"{a}cklund transformation for the coupled evolution equations (\ref{uequation})-(\ref{vequation}).


The above Douboux transformation is for the general coupled system (\ref{uequation})-(\ref{vequation}). Now we consider
the reduction of this DT for the nonlocal NLS equation (\ref{e:PTNLS}). Due to the potential symmetry (\ref{QS}), it can be shown by direct calculations that the matrix $U$  possesses the following symmetry,
\begin{eqnarray}
\sigma_{1} U^*(-x,t,-\lambda^*) \sigma_{1}= - U(x, t, \lambda). \label{UVS2}
\end{eqnarray}
Using this symmetry and the zero-curvature equation (\ref{zero_curvature}), we can derive the corresponding symmetry of the matrix $V$ by utilizing the fact that, for the given matrix $U$ in (\ref{Uform}), the matrix $V$ which satisfies the zero-curvature equation (\ref{zero_curvature}) with the specific form of $\lambda$-dependence as in Eq. (\ref{Vform}) is unique. This $V$ symmetry is
\begin{eqnarray}
\sigma_{1}V^{*}(-x,t,-\lambda^*)\sigma_{1}=V(x,t,\lambda).
\end{eqnarray}

Using these $U$ and $V$ symmetries, we can derive the symmetries of wave functions $\Phi$ and adjoint wave functions $\Psi$, and hence the symmetry of the Darboux transformation for the nonlocal NLS equation (\ref{e:PTNLS}). Applying these symmetries to the spectral problems (\ref{Laxpairxp})-(\ref{Laxpairtp}), we get
\begin{eqnarray}\label{WFreduction}
\left[\sigma_{1}\Phi^*(-x)\right]_{x}=U(x,-\lambda^*)\left[\sigma_{1}\Phi^*(-x)\right],
\end{eqnarray}
and
\begin{eqnarray}\label{WFreduction2}
\left[\sigma_{1}\Phi^*(-x)\right]_{t}=V(x,-\lambda^*)\left[\sigma_{1}\Phi^*(-x)\right].
\end{eqnarray}
Thus, if $\Phi(x)$ is a wave function of the linear system (\ref{Laxpairxp})-(\ref{Laxpairtp}) at $\lambda$, then $\sigma_{1}\Phi^*(-x)$ is a wave function of this same system at $-\lambda^*$. This symmetry has been reported in \cite{AblowitzMussNonli2016}. In the same way, we find that if $\Psi(x)$ is an adjoint wave function of the adjoint linear system (\ref{AdLaxpairxp})-(\ref{AdLaxpairtp}) at $\lambda$, then $\Psi^*(-x)\sigma_{1}$ is an adjoint wave function of the adjoint system at $-\lambda^*$.

From the above symmetries, we see that if $\lambda_1$ and $\zeta_1$ are purely imaginary, i.e.,
$\lambda_1, \zeta_1 \in \textrm{i} \mathbb{R}$, then functions $\sigma_{1}\Phi_1^*(-x)$, $\Phi_1(x)$ would satisfy the same Lax pair equations (\ref{Laxpairxp})-(\ref{Laxpairtp}) at $\lambda=\lambda_1$, and functions $\Psi_1^*(-x)\sigma_{1}$, $\Psi_1(x)$ would satisfy the same adjoint Lax pair equations (\ref{AdLaxpairxp})-(\ref{AdLaxpairtp}) at $\lambda=\zeta_1$. In this case, if $\sigma_{1}\Phi_1^*(-x)$ and $\Phi_1(x)$ are linearly dependent on each other, and $\Psi_1^*(-x)\sigma_{1}$, $\Psi_1(x)$ are linearly dependent on each other, then the Darboux transformation (\ref{eDT}) would preserve the potential reduction (\ref{qS}) and thus be a Darboux transformation for the nonlocal NLS equation (\ref{e:PTNLS}). Specifically, we have the following result.

\vspace{0.1cm}

\textbf{Proposition 1}. If $\lambda_{1}, \zeta_{1} \in \textmd{i}\mathbb{R}$,
\[ \label{PhiPsicond}
\sigma_{1}\Phi_1^*(-x)=\alpha \hspace{0.05cm} \Phi_1(x), \quad \Psi_1^*(-x)\sigma_{1}=\beta \hspace{0.05cm} \Psi_1(x),
\]
where $\alpha$ and $\beta$ are complex constants, then the Darboux matrix (\ref{eDT}) is a Darboux transformation for
the focusing nonlocal NLS equation (\ref{e:PTNLS}).

This proposition can be readily proved by checking that the new potential matrix $Q_{[1]}$ from Eq. (\ref{DTP}) satisfies the symmetry (\ref{qS}) under conditions (\ref{PhiPsicond}).

\vspace{0.2cm}
\textbf{Remark 2.} Under conditions (\ref{PhiPsicond}), it is easy to show that $|\alpha|=|\beta|=1$.

\vspace{0.2cm}
\textbf{Remark 3.} A similar but more restrictive proposition was presented in \cite{HXLM2016}, where $\alpha, \beta$ were required to be $\pm 1$. Our proposition above gives a more general DT for $\lambda_{1}, \zeta_{1} \in \textmd{i}\mathbb{R}$.

\vspace{0.1cm}
\textbf{Remark 4.} If $\lambda_{1}, \zeta_{1} \notin \textmd{i}\mathbb{R}$, then another DT reduction exists for the focusing nonlocal NLS equation (\ref{e:PTNLS}) \cite{HXLM2016}. However, this DT is irrelevant to our rogue wave calculations.


\vspace{0.2cm}
The $N$-fold Darboux transformation is a $N$ times iteration of the elementary Darboux transformation. These $N$ iterations of the elementary DT can be lumped together into a single $N$-fold Darboux matrix, which would yield a concise algebraic expression for the new solutions. For rogue-wave calculations, the relevant $N$-fold Darboux matrix is given in the following proposition.

\vspace{0.2cm}
\textbf{Proposition 2}.  \emph{The $N$-fold Darboux transformation matrix for the focusing nonlocal NLS equation (\ref{e:PTNLS}) can be represented as}
\begin{eqnarray}\label{DT}
  T_{N}=I- Y M^{-1}D^{-1} X,
\end{eqnarray}
\emph{where},
\begin{eqnarray*}\nonumber
&&Y=\left[\  \Phi_{1},\ \Phi_{2}, \ \ldots, \ \Phi_{N} \ \right]_{2\times N}, \quad X=\left[\  \begin{array}{c} \Psi_{1} \\
 \Psi_{2} \\
 \vdots \\ \Psi_{N}  \end{array} \right]_{N\times 2},\\
&&M=\left(m^{(N)}_{i,j}\right)_{1\leq i,j \leq N}, \quad  m^{(N)}_{i,j}=\frac{\Psi_{i} \Phi_{j} }{\lambda_{j}-\zeta_{i}},\\
&&D=\textrm{diag}\left( \lambda-\zeta_{1},  \lambda-\zeta_{2},\ldots,  \lambda-\zeta_{N}\right),
\end{eqnarray*}
$\lambda_{k}, \zeta_{k} \in \textmd{i}\mathbb{R}$,
$\Phi_{k}\equiv \Phi(x,t,\lambda_{k})$ solves the spectral equation (\ref{Laxpairxp})-(\ref{Laxpairtp}) at $\lambda=\lambda_{k}$, $\Psi_{k}\equiv \Psi(x,t,\zeta_{k})$ solves the adjoint spectral equation (\ref{AdLaxpairxp})-(\ref{AdLaxpairtp}) at $\lambda=\zeta_{k}$,
$$
\sigma_{1}\Phi^*(-x,t, \lambda_k)=\alpha_k \hspace{0.05cm} \Phi(x,t, \lambda_k),
\quad \Psi^*(-x, t, \zeta_k)\sigma_{1}=\beta_k \hspace{0.05cm} \Psi(x,t,\zeta_k),
$$
and $|\alpha_k|=|\beta_k|=1$. \emph{Moreover, the B\"{a}cklund transformation between potential functions is:}
\begin{eqnarray}\label{gDTpotential}
u^{[N]}=u- \textrm{i}\left[\sigma_{3}, \sum_{i,j=1}^{N}\Phi_{i} (M^{-1})_{i,j} \Psi_{j}\right]_{1,2}=
u+2\textmd{i}
\frac{\left|
\begin{array}{cc}
 M & X_{2} \\
 Y_{1} & 0 \\
\end{array}
\right|}
{\left| M \right|},
\end{eqnarray}
\emph{where $Y_1$ represents the first row of matrix $Y$, and $X_{2}$ represents the second column of matrix $X$.}

\vspace{0.2cm}
This $N$-fold DT has been reported in \cite{HXLM2016}, and the last expression in Eq. (\ref{gDTpotential}) can be found in \cite{Yang2010}. The proof of this theorem can be given along the lines of \cite{Yang2010,BGLM2015}.

\subsection{Derivation of rogue waves} \label{e:section32}
To derive general formulas for rogue waves, we first need the general eigenfunctions solved from the linear system
(\ref{Laxpairxp})-(\ref{Laxpairtp}) and its adjoint system (\ref{AdLaxpairxp})-(\ref{AdLaxpairtp}). Choosing a plane wave solution $u_{[0]}=e^{-2 \textmd{i} t}$ to be the seed solution and introducing a diagonal matrix $\mathcal{D}=\mbox{diag}\left(e^{-\textmd{i}t}, e^{\textmd{i}t}\right)$, we can derive a general wave function for the linear system
(\ref{Laxpairxp})-(\ref{Laxpairtp}) as
\[ \label{Spectraleigen0}
\Phi(x,t)=\mathcal{D}\phi(x,t),
\]
where
\begin{eqnarray}\label{Spectraleigen}
&&\phi(x,t)=\left(
  \begin{array}{c}
    c_{1}e^A+c_{2}e^{-A}  \\
    c_{4}e^A+c_{3}e^{-A} \\
  \end{array}
\right), \\
&& A=\sqrt{-\lambda^2-1}(x-2 \lambda t+\textrm{i}\theta), \nonumber \\
&& c_{3}=c_{2} \left(\textrm{i} \lambda-\sqrt{-\lambda^2-1}\right), \quad c_{4}= c_{1}\left(\textrm{i} \lambda+\sqrt{-\lambda^2-1}\right),  \nonumber
\end{eqnarray}
and $c_{1}$, $c_{2}$, $\theta$ are arbitrary complex constants. Imposing the conditions of
$\lambda$ being purely imaginary, $|\lambda|>1$, $\theta$ real, and
\begin{eqnarray}\label{Condition12}
\alpha c_{2}=   c^*_{1} \left( \textrm{i} \lambda+\sqrt{-\lambda^2-1} \right), \quad |\alpha|=1,
\end{eqnarray}
the above wave function would satisfy the symmetry condition (\ref{PhiPsicond}). Through a simple Gauge (phase) transformation on the complex constant $c_1$, we can normalize $\alpha=1$.

Similarly, for the same seed solution $u_{[0]}=e^{-2 \textmd{i} t}$, we can derive the adjoint wave functions to Eqs. (\ref{AdLaxpairxp})-(\ref{AdLaxpairtp}) satisfying the symmetry condition (\ref{PhiPsicond}) as
\[
\Psi(x,t)=\psi(x,t)\mathcal{D}^{*},
\]
where
\begin{eqnarray}\label{Adeigen}
&&\psi(x,t)=
\left(\widehat{c}_{1}e^{\widehat{A}}+\widehat{c}_{2}e^{-\widehat{A}}, \quad     \widehat{c}_{4}e^{\widehat{A}}+\widehat{c}_{3}e^{-\widehat{A}} \right), \\
&& \widehat{A}=\sqrt{-\lambda^2-1}(x-2 \lambda t+\textrm{i}\widehat{\theta}), \nonumber \\
&& \widehat{c}_{3}=\widehat{c}_{2} \left(-\textrm{i} \lambda-\sqrt{-\lambda^2-1}\right), \quad \widehat{c}_{4}= \widehat{c}_{1}\left(-\textrm{i} \lambda+\sqrt{-\lambda^2-1}\right),  \nonumber
\end{eqnarray}
$\lambda$ is purely imaginary, $|\lambda|>1$, $\widehat{\theta}$ is real, and
\begin{eqnarray}
\beta \widehat{c}_{2}=   \widehat{c}^*_{1} \left( -\textrm{i} \lambda+\sqrt{-\lambda^2-1} \right), \quad |\beta|=1.
\end{eqnarray}
Through a Gauge transformation on the complex constant $\widehat{c}_1$,  we can normalize $\beta=1$. It is noted that this $\Psi(x,t)$ solution can also be derived directly from the above $\Phi(x,t)$ solution using the fact that, for the present $x$-independent
seed solution $u_{[0]}$, if $\Phi(x,t, \zeta)$ solves the original linear system (\ref{Laxpairxp})-(\ref{Laxpairtp}), then
$\Psi(x,t, \lambda)=\Phi^\dagger(x,t, \zeta^*)$ solves the adjoint system (\ref{AdLaxpairxp})-(\ref{AdLaxpairtp}).

Due to the free complex constants $c_1$ and $\widehat{c}_1$, we get two types of wave functions and adjoint wave functions. When $c_1$ is taken as a real constant and denoting $\lambda=ih$ with $h>1$,  then after a constant normalization, the wave function $\phi(x,t)$ becomes
\begin{eqnarray} \label{e:phisinh}
&& \phi(x,t,\lambda)= \frac{1}{\sqrt{h-1}} \left(
                    \begin{array}{c}
 \sinh\left[A+ \frac{1}{2}\ln\left(h+\sqrt{h^2-1}\right)\right] \\
 \sinh \left[-A+ \frac{1}{2}\ln\left(h+\sqrt{h^2-1}\right)\right]  \\
                    \end{array}
                  \right), \\
&&A=\sqrt{h^2-1} (x-2 \textrm{i} h t+\textrm{i}\theta),    \label{Eq:A}
\end{eqnarray}
where $\theta$ is a real constant. Here, the scaling constant $1/\sqrt{h-1}$ is introduced so that this wave function does not approach zero in the limit of $h\to 1$ (i.e., $\lambda\to \textmd{i}$). This scaling of the wave function clearly does not affect the solution in view of Proposition 2. Hereafter, we call this wave function type-a.

When $c_1$ is taken as a purely imaginary constant, then after a constant normalization, the wave function $\phi(x,t)$ becomes
\begin{eqnarray} \label{e:phicosh}
\phi(x,t,\lambda)=\left(
                    \begin{array}{c}
\cosh \left[ A+ \frac{1}{2}\ln\left(h+\sqrt{h^2-1}\right) \right]\\
 -\cosh \left[A- \frac{1}{2}\ln\left(h+\sqrt{h^2-1}\right)\right] \\
                    \end{array}
                  \right),
\end{eqnarray}
where $A$ is the same as that in (\ref{Eq:A}). Hereafter, we call this wave function type-b.

If we take the limit of $h\to 1^+$ in the above solution (\ref{e:phicosh}), we also get a special wave function
\[  \label{e:phiconstant}
\phi_0=\left(
                    \begin{array}{c}  1 \\ -1
                     \end{array}
                  \right)
\]
at the spectral parameter $\lambda=\textmd{i}$. We call this wave function type-c.

Similarly, we get several types of adjoint wave functions. If $\widehat{c}_1$ is taken as a real constant and denoting $\lambda=-\textmd{i}\widehat{h}$ with $\widehat{h}>1$, then after a constant normalization, the adjoint wave function (\ref{Adeigen}) becomes
\begin{eqnarray} \label{e:psisinh}
&& \psi^T(x,t,\lambda)= \frac{1}{\sqrt{\widehat{h}-1}}\left(
                    \begin{array}{c}
 \sinh \left[ \widehat{A}+\frac{1}{2}\ln\left(\widehat{h}+\sqrt{\widehat{h}^2-1}\right)\right]\\
 \sinh \left[ -\widehat{A}+\frac{1}{2}\ln\left(\widehat{h}+\sqrt{\widehat{h}^2-1}\right) \right] \\
                    \end{array}
                  \right),\\ \nonumber
&&\widehat{A}=\sqrt{\widehat{h}^2-1} (x+2 \textrm{i} \widehat{h} t+\textrm{i}\widehat{\theta}),
\end{eqnarray}
where $\widehat{\theta}$ is a real constant. Since this adjoint wave function is the counterpart of type-a wave function, we call it of type-a as well. If $\widehat{c}_1$ is taken as a purely imaginary constant, then after a constant normalization, the adjoint wave function (\ref{Adeigen}) becomes
\begin{eqnarray} \label{e:psicosh}
&&  \psi^T(x,t,\lambda)=\left(
                    \begin{array}{c}
\cosh \left[ \widehat{A}+ \frac{1}{2}\ln\left(\widehat{h}+\sqrt{\widehat{h}^2-1}\right) \right]\\
-\cosh \left[\widehat{A}- \frac{1}{2}\ln\left(\widehat{h}+\sqrt{\widehat{h}^2-1}\right)\right] \\
                    \end{array}
                  \right).
\end{eqnarray}
This adjoint wave function will be called type-b since it is the counterpart of the type-b wave function.
In the limit of $\widehat{h}\to 1^+$, this latter adjoint wave function reduces to
\[  \label{e:psiconstant}
\psi_0=(1, -1)
\]
at the spectral parameter $\lambda=-\textmd{i}$. It will be called type-c.

Rogue waves are rational solutions. To derive rogue waves from the above wave functions and adjoint wave functions using Proposition 2, we need to choose spectral parameters $\lambda_{k}$ and $\zeta_{k}$ so that the exponents $A$ and $\widehat{A}$ vanish themselves or vanish under certain limits. These exponents would vanish when $\lambda=\pm \textmd{i}$. Thus, we can take $\lambda_{k}$ to be $\textmd{i}$ or approach $\textmd{i}$, and take $\zeta_{k}$ to be $-\textmd{i}$ or approach $-\textmd{i}$. Since wave functions and adjoint wave functions in the Darboux transformation for the nonlocal NLS equation (\ref{e:PTNLS}) are unrelated (see Proposition 2), and several different types of wave functions and adjoint wave functions exist, we have a lot of freedom in the construction of solutions. Choices of different types of wave functions and adjoint wave functions will lead to different types of rogue waves. This will be treated separately below.

1. We choose the wave functions to be all type-a with spectral parameters $\lambda=\lambda_k$ and constant $\theta=\theta_k$, where
\[ \label{e:parametersI}
\lambda_k=\textmd{i}(1+\epsilon_k^2), \quad \theta_k=\sum_{j=0}^{n-1}s_{j}\epsilon_{k}^{2j},  \quad 1\le k\le n,
\]
and $s_0, s_1, \dots, s_{n-1}$ are real constants. It is important that these real constants $s_j$ are $k$-independent.
In addition, we choose the adjoint wave functions to be also all type-a with spectral parameters $\lambda=\zeta_k$ and constant $\widehat{\theta}=\widehat{\theta}_k$, where
\[ \label{e:parametersII}
\zeta_k=-\textmd{i}(1+\widetilde{\epsilon}_k^2), \quad \widehat{\theta}_k=\sum_{j=0}^{n-1}r_{j}\widetilde{\epsilon}_{k}^{2j}, \quad  1\le k\le n,
\]
and $r_0, r_1, \dots, r_{n-1}$ are $k$-independent real constants. Notice that the wave function $\phi(x,t,\lambda)$ in (\ref{e:phisinh}) with $\lambda=\textmd{i}(1+\epsilon^2)$ and adjoint wave function $\psi(x,t,\zeta)$ in (\ref{e:psisinh}) with $\zeta=-\textmd{i}(1+\epsilon^2)$ are both even functions of $\epsilon$. Thus, we can expand
\[
\phi(x,t,\lambda)=\sum_{k=0}^{\infty} \phi^{(k)} \epsilon^{2k}, \quad
\psi(x,t,\zeta)=\sum_{k=0}^{\infty} \psi^{(k)} \widetilde{\epsilon}^{2k},
\]
\[
\frac{\psi(x,t,\zeta)\phi(x,t,\lambda)}{\lambda-\zeta}=
\sum_{k=0}^{\infty}\sum_{l=0}^{\infty}m_{k,\hspace{0.04cm} l}\hspace{0.04cm}
\widetilde{\epsilon}^{2k} \epsilon^{2l},
\]
where
\[
\phi^{(k)}=\lim_{\epsilon\rightarrow 0}\frac{\partial^{2k}\phi(x,t,\lambda)}{(2k)! \partial \epsilon^{2k}},
\quad \psi^{(k)}=\lim_{\widetilde{\epsilon}\rightarrow 0}\frac{\partial^{2k}\psi(x,t,\zeta)}{(2k)! \partial \widetilde{\epsilon}^{2k}},
\]
and
\[
m_{k,l}=\lim_{\epsilon, \widetilde{\epsilon} \rightarrow 0}\frac{1}{(2k-2)!(2l-2)!}\frac{\partial^{2k+2l-4}}{\partial \widetilde{\epsilon}^{2k-2}\partial \epsilon^{2l-2}}
\left[\frac{\psi(x,t,\zeta)\phi(x,t,\lambda)}{\lambda-\zeta}\right].
\]
Applying these expansions to each matrix element in the B\"acklund transformation (\ref{gDTpotential}), performing simple determinant manipulations and taking the limits of $\epsilon_{k},  \widetilde{\epsilon}_{k} \rightarrow 0$ ($1\le k\le n$), we derive the type-I rogue waves as given in Theorem 1.

2. We choose the wave functions to be all type-a as in the first case, but choose the adjoint wave functions to be a mixture of type-a and type-c. More specifically, we choose the first adjoint wave function $\Psi_1$ to be type-c and the other adjoint wave functions to be type-a, i.e., $\Psi_1=\psi_0 \mathcal{D}^{*}$ at spectral parameter $\lambda=-\textmd{i}$, and $\Psi_k=\psi(x,t,\zeta_k)\mathcal{D}^{*} (2\le k\le n)$, where $\psi(x,t,\zeta_k)$ is the type-a adjoint wave function (\ref{e:psisinh}) at spectral parameters $\lambda=\zeta_k$ and constant $\widehat{\theta}=\widehat{\theta}_k$ as in Eq. (\ref{e:parametersII}). Then, repeating the limit process as in the previous case, we derive the type-II rogue waves as given in Theorem 2.

3. We choose the wave functions to be all type-a as in the first two cases, but choose the adjoint wave functions to be all type-b, with spectral parameters $\lambda=\zeta_k$ and constant $\widehat{\theta}=\widehat{\theta}_k$ as in Eq. (\ref{e:parametersII}). Then repeating the same limit processes as in the first two cases, we derive the type-III rogue waves as given in Theorem 3.

\subsection{Algebraic expressions of rogue waves}
In this subsection, we prove Theorem 4, which gives more explicit algebraic expressions for the rogue waves in Theorems 1-3.

The basic idea is to notice that, the matrix elements in Theorems 1-3, which are derivatives of certain functions in the small $\epsilon$ and $\widetilde{\epsilon}$ limits, are nothing but the coefficients of power expansions of those functions in $\epsilon$ and $\widetilde{\epsilon}$. Thus, we need to derive these power expansions.

For this purpose, we first recall the following three expansions,
\begin{eqnarray}
&& \sqrt{2+\epsilon^2}=\sum_{k=0}^{\infty}p_{k}\epsilon^{2k},   \label{e:expansion1} \\
&& \ln\left(1+\frac{\epsilon^2}{2}\right)=\sum_{k=1}^{\infty}q_{k}\epsilon^{2k},  \label{e:expansion2} \\
&&  \frac{1}{2}\ln\left(1+\epsilon^2+\epsilon\sqrt{2+\epsilon^2}\right)=\sum_{k=0}^{\infty}w_k\epsilon^{2k+1}, \label{e:expansion3}
\end{eqnarray}
where
\begin{eqnarray*}
&&  p_k=\sqrt{2}{\frac{1}{2} \choose k}\left(\frac{1}{2}\right)^k,\quad  {\frac{1}{2} \choose k}=\frac{\frac{1}{2}( \frac{1}{2}-1)\cdots(\frac{1}{2}-k+1)}{k!},\\
&&  q_{k}=\frac{(-1)^{k+1}}{k\cdot 2^k},\quad
w_k=\frac{\sqrt{2}(2k)!}{2^{3k+1}(k!)^2}\frac{\left(-1\right)^{k}}{(2k+1)}.
\end{eqnarray*}
The first two expansions are straightforward. Regarding the third expansion (\ref{e:expansion3}), notice that the function on its left side is odd in $\epsilon$; hence its expansion does not contain any even powers of $\epsilon$. To prove this expansion, we notice that
\begin{eqnarray*}
&& \frac{1}{2}\ln\left(1+\epsilon^2+\epsilon\sqrt{2+\epsilon^2}\right)=\ln\left(\frac{\epsilon}{\sqrt{2}}+\sqrt{1+\left(\frac{\epsilon}{\sqrt{2}}\right)^2}\right)
=\textmd{i}\hspace{0.05cm}\mbox{arcsin}\left(\frac{-\textmd{i}\epsilon}{\sqrt{2}}\right),
\end{eqnarray*}
because $\mbox{arcsin}(x)=-\textmd{i}\ln\left(\textmd{i}x+\sqrt{1-x^2}\right)$. Then, using the Taylor expansion of $\mbox{arcsin}(x)$, the third expansion (\ref{e:expansion3}) can then be proved.

Next, we define
\[
\Theta^{\pm}(\epsilon)=\pm A+ \frac{1}{2}\ln\left(h+\sqrt{h^2-1}\right),
\]
\[
\widetilde{\Theta}^{\pm}(\widetilde{\epsilon})=\pm\widehat{A}+\frac{1}{2}\ln\left(\widehat{h}+\sqrt{\widehat{h}^2-1}\right),
\]
where $A, h, \widehat{A}$ and $\widehat{h}$ are as given in Theorem 1. These functions appeared in the exponents of the wave functions and adjoint wave functions in Theorems 1-3; thus their power expansions are needed.

Using the expansions (\ref{e:expansion1}) and (\ref{e:expansion3}), we can calculate the expansion for $\Theta^{+}(\epsilon)$ as
\begin{eqnarray*}
&&  \Theta^{+}(\epsilon)=\epsilon \sqrt{2+\epsilon^2}\left(  x-2 \textrm{i} (1+\epsilon^2)t+ \sum_{k=0}^{\infty} \textrm{i}s_{k} \epsilon^{2k} \right)
+\frac{1}{2} \ln\left(1+\epsilon^2+\epsilon\sqrt{2+\epsilon^2}\right)\\
&&  \hspace{0.72cm}= \epsilon \left( \sum_{k=0}^{\infty}p_k\epsilon^{2k}\right)\left( \sum_{k=0}^{\infty}\left(\delta_{k,0} x - 2\textrm{i}t  + \textrm{i}s_{k} \right)\epsilon^{2k}\right)+
\sum_{k=0}^{\infty}w_k\epsilon^{2k+1},
\end{eqnarray*}
where $\delta_{k,0}$ is the Kronecker delta notation. Thus,
\[  \label{e:expandTheta+}
\Theta^{+}(\epsilon)= \sum_{k=0}^{\infty}X^{+}_{2k+1} \epsilon^{2k+1},
\]
where $X_{2k+1}^{+}$ is as given in Theorem 4.
Similarly, we obtain the expansion for $\widetilde{\Theta}^{+}(\widetilde{\epsilon})$ as
\begin{eqnarray}
&& \widetilde{\Theta}^{+}(\widetilde{\epsilon})=\sum_{k=0}^{\infty}\widetilde{X}^{+}_{2k+1} \widetilde{\epsilon}^{\ 2k+1},
\end{eqnarray}
where $\widetilde{X}^{+}_{2k+1}$ is as shown in Theorem 4.

Regarding $\Theta^{-}(\epsilon)$ and $\widetilde{\Theta}^{-}(\widetilde{\epsilon})$, we notice that
\begin{eqnarray} \label{ncsymmetry}
\Theta^{-}(x)=[\Theta^{+}(-x)]^{*}, \quad
\widetilde{\Theta}^{-}(x)=[\widetilde{\Theta}^{+}(-x)]^*.
\end{eqnarray}
From these relations, we get the expansions for $\Theta^{-}$ and  $\widetilde{\Theta}^{-}$ as
\[
\Theta^{-}(\epsilon)= \sum_{k=0}^{\infty}X^{-}_{2k+1} \epsilon^{2k+1}, \quad
\widetilde{\Theta}^{-}(\widetilde{\epsilon})= \sum_{k=0}^{\infty}\widetilde{X}^{-}_{2k+1} \widetilde{\epsilon}^{2k+1},
\]
where $X^{-}_{2k+1}$ and $\widetilde{X}^{-}_{2k+1}$ are as defined in Theorem 4.

From these expansions, we see that $\Theta^{\pm}(\epsilon)$ and $\widetilde{\Theta}^{\pm}(\widetilde{\epsilon})$ are odd functions. Thus,
\[
-\Theta^{\pm}(\epsilon)=\Theta^{\pm}(-\epsilon), \quad
-\widetilde{\Theta}^{\pm}(\widetilde{\epsilon})=\widetilde{\Theta}^{\pm}(-\widetilde{\epsilon}).
\]
Using these relations, functions $-\Theta^{\pm}(\epsilon)$ and $-\widetilde{\Theta}^{\pm}(\widetilde{\epsilon})$ can be expanded as
\[
-\Theta^{\pm}(\epsilon)=\sum_{k=0}^{\infty}X^{\pm}_{2k+1} \left(-\epsilon\right)^{2k+1}, \quad
-\widetilde{\Theta}^{\pm}(\widetilde{\epsilon})=\sum_{k=0}^{\infty}\widetilde{X}^{\pm}_{2k+1} \left(-\widetilde{\epsilon}\right)^{2k+1}.
\]

Next, we also need to expand $1/(\lambda-\zeta)$, where $\lambda=\textrm{i}(1+\epsilon^2)$ and $\zeta=-\textrm{i}(1+ \widetilde{\epsilon}^{\hspace{0.075cm} 2})$. This expansion can be obtained as follows,
\begin{eqnarray}
&& \frac{1}{\lambda-\zeta} =
\frac{-2 \textrm{i}}{(2+\epsilon^2)(2+\tilde{\epsilon}^{\hspace{0.05cm}2})}
\left(1-\frac{\epsilon^2 \tilde{\epsilon}^2}{(2+\epsilon^2)(2+\tilde{\epsilon}^{2})}\right)^{-1}   \nonumber \\
&&  \hspace{1.25cm}= \left(\frac{1}{2 \textrm{i}}\right)
\frac{1}{(1+ \frac{\epsilon^2}{2})(1+ \frac{\widetilde{\epsilon}^2}{2})}
\sum_{\nu=0}^{\infty} \left( \frac{\epsilon^2 \tilde{\epsilon}^2}{(2+\epsilon^2)(2+\tilde{\epsilon}^{2})} \right)^\nu  \nonumber \\
&&\hspace{1.25cm} =\left(\frac{1}{2 \textrm{i}}\right) \sum_{\nu=0}^{\infty} \left( \frac{\epsilon^{2} \tilde{\epsilon}^{ 2}}{4} \right)^{\nu}
\exp\left[-(\nu+1)\ln\left(1+\frac{\epsilon^2}{2}\right)-(\nu+1)\ln\left(1+\frac{\widetilde{\epsilon}^2}{2}\right)\right] \nonumber \\
&& \hspace{1.25cm} =\left(\frac{1}{2 \textrm{i}}\right) \sum_{\nu=0}^{\infty} \left( \frac{\epsilon^{2} \tilde{\epsilon}^{ 2}}{4} \right)^{\nu} \exp\left( \sum_{k=1}^{\infty} X_{2k}^{\pm}(\nu) \epsilon^{2k} + \sum_{k=1}^{\infty} \widetilde{X}_{2k}^{\pm}(\nu) \widetilde{\epsilon}^{\ 2k}\right),  \label{e:expandlambdazeta}
\end{eqnarray}
where $X_{2k}^{\pm}(\nu)$ and $\widetilde{X}_{2k}^{\pm}(\nu)$ are as given in Theorem 4.

As in Theorem 4, we introduce vectors $\textbf{X}^{\pm}$ and $\widetilde{\textbf{X}}^{\pm}$. Then, using Schur polynomials defined in section 2, we have
\begin{eqnarray} \label{e:Schurexpand}
\exp\left(\sum_{k=0}^{\infty} {X}^{\pm}_{k}\epsilon^{k}\right)=\sum_{k=0}^{\infty}S_{k}(\textbf{X}^{\pm})\epsilon^{k},\
\exp\left(\sum_{k=0}^{\infty} \widetilde{X}^{\pm}_{k} \widetilde{\epsilon}^{\hspace{0.06cm}k}\right)=\sum_{k=0}^{\infty}S_{k}(\widetilde{\textbf{X}}^{\pm})\widetilde{\epsilon}^{\hspace{0.06cm}k}.
\end{eqnarray}

Now, we are ready to derive all the matrix elements in Theorems 1-3 in terms of purely algebraic expressions.
First, we consider type-I rogue-wave solutions in Theorem 1. Its matrix elements $m_{i,j}^{(1)}$ defined in (\ref{detmij1}) come from the $\mathcal{O}(\epsilon^{2i-2}\widetilde{\epsilon}^{\hspace{0.02cm }2j-2})$ coefficients in the two-variables Taylor expansion of the function
\begin{eqnarray}\label{2variableTaylor}
\frac{\psi(\zeta)\phi(\lambda)}{\lambda-\zeta}=\frac{\psi_{1}(\zeta)\phi_{1}(\lambda)}{\lambda-\zeta}+
\frac{\psi_{2}(\zeta)\phi_{2}(\lambda)}{\lambda-\zeta},
\end{eqnarray}
where $\phi=(\phi_1, \phi_2)^T$ and $\psi=(\psi_1, \psi_2)$ are given in Eqs. (\ref{Eigenfunction})-(\ref{AdEigenfunction}) with
 $\lambda=\textmd{i}(1+\epsilon^2)$ and $\zeta=-\textmd{i}(1+\widetilde{\epsilon}^2)$.

By using expansions (\ref{e:expandTheta+})-(\ref{e:Schurexpand}), a direct calculation shows that
\begin{eqnarray*}
&& \frac{\psi_{1}(\zeta)\phi_{1}(\lambda)}{\lambda-\zeta}=\left( \frac{1}{2 \epsilon\widetilde{\epsilon}} \right)  \frac{\cosh\left[\Theta^{+}(\epsilon)+\widetilde{\Theta}^{+}(\widetilde{\epsilon})\right]-
\cosh\left[\Theta^{+}(\epsilon)-\widetilde{\Theta}^{+}(\widetilde{\epsilon})\right]}{\lambda-\zeta}\\
&&\hspace{2.1cm}=\frac{1}{2 \textrm{i}}  \sum_{\nu=0}^{\infty} \left( \frac{\epsilon^{2} \tilde{\epsilon}^{ 2}}{4} \right)^{\nu}
\sum_{k=0}^{\infty}S_{2k+1}(\textbf{X}^{+})\epsilon^{2k} \sum_{l=0}^{\infty}S_{2l+1}(\widetilde{\textbf{X}}^+)\widetilde{\epsilon}^{2l}\\
&&  \hspace{2.1cm} =   \frac{1}{2 \textrm{i}} \sum_{i=0}^{\infty}\sum_{j=0}^{\infty}\sum_{\nu=0}^{\min\{i,j\}} \frac{1}{4^{\nu}}S_{2i-2\nu+1}(\textbf{X}^{+})
S_{2j-2\nu+1}(\widetilde{\textbf{X}}^{+})\epsilon^{2i} \widetilde{\epsilon}^{\hspace{0.05cm}2j},\\
&& \frac{\psi_{2}(\zeta)\phi_{2}(\lambda)}{\lambda-\zeta}=\left( \frac{1}{2 \epsilon\widetilde{\epsilon}} \right)  \frac{\cosh\left[\Theta^{-}(\epsilon)+\widetilde{\Theta}^{-}(\widetilde{\epsilon})\right]-
 \cosh\left[\Theta^{-}(\epsilon)-\widetilde{\Theta}^{-}(\widetilde{\epsilon})\right]}{\lambda-\zeta}\\
 &&\hspace{2.1cm}=\frac{1}{2 \textrm{i}}  \sum_{\nu=0}^{\infty} \left( \frac{\epsilon^{2} \tilde{\epsilon}^{ 2}}{4} \right)^{\nu}
\sum_{k=0}^{\infty}S_{2k+1}(\textbf{X}^{-})\epsilon^{2k} \sum_{l=0}^{\infty}S_{2l+1}(\widetilde{\textbf{X}}^-)\widetilde{\epsilon}^{2l}\\
&&  \hspace{2.1cm} = \frac{1}{2 \textrm{i}} \sum_{i=0}^{\infty}\sum_{j=0}^{\infty}\sum_{\nu=0}^{\min\{i,j\}} \frac{1}{4^{\nu}}S_{2i-2\nu+1}(\textbf{X}^{-})
S_{2j-2\nu+1}(\widetilde{\textbf{X}}^{-})\epsilon^{2i} \widetilde{\epsilon}^{\hspace{0.05cm}2j}.
\end{eqnarray*}
Inserting these two parts into (\ref{2variableTaylor}), the coefficient $m_{i,j}^{(1)}$ of the
$\mathcal{O}(\epsilon^{2i-2} \widetilde{\epsilon}^{\hspace{0.02cm }2j-2})$ term is then
\begin{eqnarray*}
m_{i,j}^{(1)}= \frac{1}{2 \hspace{0.02cm} \textrm{i}} \sum_{\nu=0}^{\min\{i-1,j-1\}}\frac{1}{4^{\nu}} \left[S_{2i-2\nu-1}(\textbf{X}^{+})
S_{2j-2\nu-1}(\widetilde{\textbf{X}}^{+}) +S_{2i-2\nu-1}(\textbf{X}^{-})
S_{2j-2\nu-1}(\widetilde{\textbf{X}}^{-}) \right],
\end{eqnarray*}
which matches that given in Theorem 4.

To express $\mu^{(1)}$ and $\nu^{(1)}$ in matrix $\tau^{(1)}_{1}$, we also need the expansions of wave functions and adjoint wave functions. Derivation of these expansions is easier. Recall that
\begin{eqnarray*}
&&  \phi_{1}(\lambda)=  \frac{ \sinh\left(\Theta^{+}\right)}{\epsilon}, \quad  \psi_{2}(\zeta)=   \frac{\sinh (\widetilde{\Theta}^{-})}{\widetilde{\epsilon}}.
\end{eqnarray*}
Then, using the earlier results and following similar calculations, we get
\begin{eqnarray*}
&&  \phi_{1}(\lambda)= \sum_{j=1}^{\infty}\mu_j^{(1)} \epsilon^{2(j-1)},\ \mu_{j}^{(1)}=S_{2j-1}(\textbf{\emph{Y}}^{+}), \\
&&  \psi_{2}(\zeta)= \sum_{j=1}^{\infty}\nu_{j}^{(1)}\widetilde{\epsilon}^{2(j-1)}, \ \nu_{j}^{(1)}= S_{2j-1}(\widetilde{\textbf{\emph{Y}}}^{-}),
\end{eqnarray*}
where $\textbf{\emph{Y}}^{+}$, $\widetilde{\textbf{\emph{Y}}}^{-}$, $\mu_j^{(1)}$ and $\nu_{j}^{(1)}$ are as defined in Theorem 4.

For type-II rogue waves in Theorem 2, the only expression we need to derive is for $m^{(2)}_{1,j}$. This is again an easier derivation, and we find that
\begin{eqnarray*}
&&   \frac{\psi_{0} \phi(\lambda)}{\lambda+  \textrm{i}} = \frac{ \phi_{2}-\phi_{1}}{\textrm{i}(2+\epsilon^2)}=
 \sum_{j=1}^{\infty} m^{(2)}_{1,j} \epsilon^{2(j-1)},\\
&&  m^{(2)}_{1,j}=\frac{1 }{2\textrm{i}} \left[S_{2j-1}(\textbf{\emph{W}}^{+})-S_{2j-1}(\textbf{\emph{W}}^{-})\right],
\end{eqnarray*}
where $\textbf{\emph{W}}^{\pm}$ are as defined in Theorem 4.

At last, to derive explicit expressions for type-III rogue waves in Theorem 3, one only needs to make some modifications to the derivation for type-I rogue waves above. In this case,
\begin{eqnarray}\label{2variableTaylorIII}
\frac{\omega(\zeta)\phi(\lambda)}{\lambda-\zeta}=\frac{\omega_{1}(\zeta)\phi_{1}(\lambda)}{\lambda-\zeta}+
\frac{\omega_{2}(\zeta)\phi_{2}(\lambda)}{\lambda-\zeta},
\end{eqnarray}
where $\omega(\zeta)$ is the adjoint wave function given in Theorem 3, and $\phi(\lambda)$ is still the wave function given in Theorem 1.
From direct calculations, we get
\begin{eqnarray*}
&& \frac{\omega_{1}(\zeta)\phi_{1}(\lambda)}{\lambda-\zeta}= \left( \frac{1}{2\epsilon } \right)   \frac{   \sinh\left[\Theta^{+}(\epsilon)+\widetilde{\Theta}^{+}(\widetilde{\epsilon})\right]+
   \sinh\left[\Theta^{+}(\epsilon)-\widetilde{\Theta}^{+}(\widetilde{\epsilon})\right]}{\lambda-\zeta}\\
&& \hspace{2.1cm} =  \frac{1}{2\textrm{i}} \sum_{\nu=0}^{\infty} \left( \frac{\epsilon^{2} \tilde{\epsilon}^{ 2}}{4} \right)^{\nu}
\sum_{k=0}^{\infty}S_{2k+1}(\textbf{X}^{+})\epsilon^{2k} \sum_{l=0}^{\infty}S_{2l}(\widetilde{\textbf{X}}^{+})\widetilde{\epsilon}^{2l}\\
&& \hspace{2.1cm}  = \frac{1}{2 \textrm{i}} \sum_{i=0}^{\infty}\sum_{j=0}^{\infty}\sum_{\nu=0}^{\min\{i,j\}} \frac{1}{4^{\nu}}S_{2i-2\nu+1}(\textbf{X}^{+})
S_{2j-2\nu}(\widetilde{\textbf{X}}^{+})\epsilon^{2i} \widetilde{\epsilon}^{\hspace{0.06cm} 2j},\\
&& \frac{\omega_{2}(\zeta)\phi_{2}(\lambda)}{\lambda-\zeta}= \left( \frac{-1}{2\epsilon } \right)   \frac{   \sinh\left[\Theta^{-}(\epsilon)+\widetilde{\Theta}^{-}(\widetilde{\epsilon})\right]+
   \sinh\left[\Theta^{-}(\epsilon)-\widetilde{\Theta}^{-}(\widetilde{\epsilon})\right]}{\lambda-\zeta}\\
&& \hspace{2.1cm} =  \frac{\textrm{i}}{2} \sum_{\nu=0}^{\infty} \left( \frac{\epsilon^{2} \tilde{\epsilon}^{ 2}}{4} \right)^{\nu}
\sum_{k=0}^{\infty}S_{2k+1}(\textbf{X}^{-})\epsilon^{2k} \sum_{l=0}^{\infty}S_{2l}(\widetilde{\textbf{X}}^{-})\widetilde{\epsilon}^{2l}\\
&& \hspace{2.1cm}  = \frac{\textrm{i}}{2} \sum_{i=0}^{\infty}\sum_{j=0}^{\infty}\sum_{\nu=0}^{\min\{i,j\}} \frac{1}{4^{\nu}}S_{2i-2\nu+1}(\textbf{X}^{-})
S_{2j-2\nu}(\widetilde{\textbf{X}}^{-})\epsilon^{2i} \widetilde{\epsilon}^{\hspace{0.06cm} 2j}.
\end{eqnarray*}
Combining these two parts,
the coefficient $m_{i,j}^{(3)}$ of the $\mathcal{O}(\epsilon^{2i-2} \widetilde{\epsilon}^{\hspace{0.02cm }2j-2})$ term is then
\begin{eqnarray*}
m_{i,j}^{(3)}=\frac{1}{2 \hspace{0.02cm} \textrm{i}} \sum_{\nu=0}^{\min\{i-1,j-1\}} \frac{1}{4^{\nu}}\left[S_{2i-2\nu-1}(\textbf{X}^{+})
S_{2j-2\nu-2}(\widetilde{\textbf{X}}^{+})-S_{2i-2\nu-1}(\textbf{X}^{-})
S_{2j-2\nu-2}(\widetilde{\textbf{X}}^{-})\right],
\end{eqnarray*}
which matches that in Theorem 4. Similarly,
\begin{eqnarray*}
&&  \omega_{2}(\zeta)= - \frac{  \cosh (\widetilde{\Theta}^{-})}{\widetilde{\epsilon}}
  =  \sum_{j=1}^{\infty} \nu_{j}^{(3)}\widetilde{\epsilon}^{2(j-1)}, \quad  \nu_{j}^{(3)}=-S_{2j-2}(\widetilde{\textbf{\emph{Y}}}^{-}).
\end{eqnarray*}
Theorem 4 is then proved.

\subsection{Boundary conditions}
In this subsection, we prove Theorem 5, which gives the boundary conditions of these rogue waves as well as the degrees of their denominator polynomials.

From the definition in (\ref{Schurpoly}), we know that the Schur polynomial has the form $S_{k}(\mathbf{x})=x_{1}^k/k!+(\textrm{lower degree terms})$. Thus, the degree of the polynomial $S_{i-2\nu}(\widetilde{\textbf{\emph{X}}}^{\pm})$ in $(x,t)$ is $(i-2\nu)$, and its leading term is given by $\pm (x+2\textrm{i}t)^{i-2\nu}/(i-2\nu)!$. In addition, the degree of the polynomial $S_{j-2\nu}(\textbf{\emph{X}}^{\pm})$ is $(j-2\nu)$, and its leading term is given by $\pm (x-2\textrm{i}t)^{j-2\nu}/(j-2\nu)!$.
Thus,
\begin{eqnarray*}
\mbox{deg}(\mathcal{A}_{2j-1,\nu})=\mbox{deg}(\mathcal{B}_{2j-1,\nu})=\mbox{deg}(\widetilde{\mathcal{A}}_{2j-1,\nu})=\mbox{deg}(\widetilde{\mathcal{B}}_{2j-1,\nu})=2j-2\nu-1, \\
\mbox{deg}(\mathcal{A}_{2j-2,\nu})=\mbox{deg}(\mathcal{B}_{2j-2,\nu})=\mbox{deg}(\widetilde{\mathcal{A}}_{2j-2,\nu})
=\mbox{deg}(\widetilde{\mathcal{B}}_{2j-2,\nu})=2j-2\nu-2,
\end{eqnarray*}
but
\begin{eqnarray*}
\mbox{deg}(\mathcal{A}_{2j-1,\nu}+\mathcal{B}_{2j-1,\nu})=\mbox{deg}(\widetilde{\mathcal{A}}_{2j-1,\nu}+\widetilde{\mathcal{B}}_{2j-1,\nu})=2j-2\nu-2, \\
\mbox{deg}(\mathcal{A}_{2j-2,\nu}-\mathcal{B}_{2j-2,\nu})=\mbox{deg}(\widetilde{\mathcal{A}}_{2j-2,\nu}-\widetilde{\mathcal{B}}_{2j-2,\nu})=2j-2\nu-3.
\end{eqnarray*}
Using these relations and performing similar calculations as in Ref. \cite{OhtaJY2012}, we can show from Theorem 4 that
\begin{eqnarray*}
&&\tau^{(1)}_{0}=c_{0}^{(1)}(x^2+4t^2)^{n(n+1)/2}+(\textrm{lower degree terms}), \\
&&\tau^{(2)}_{0}=c_{0}^{(2)}(x+2\textrm{i}t)^{(n-1)(n-2)/2}(x-2\textrm{i}t)^{n(n+1)/2}+(\textrm{lower degree terms}), \\
&&\tau^{(3)}_{0}=c_{0}^{(3)}(x+2\textrm{i}t)^{n(n-1)/2}(x-2\textrm{i}t)^{n(n+1)/2}+(\textrm{lower degree terms}),
\end{eqnarray*}
where $c_0^{(1)}$, $c_0^{(2)}$ and $c_0^{(3)}$ are $n$-dependent constants. Thus, the polynomial degrees of $\tau^{(k)}_{0}$ in Theorem 5 are proved.

In addition, we find that when $n$ is odd,
\begin{eqnarray*}
&&\tau^{(1)}_{1}=ic_{0}^{(1)}(x^2+4t^2)^{n(n+1)/2}+(\textrm{lower degree terms}), \\
&&\tau^{(2)}_{1}=ic_{0}^{(2)}(x+2\textrm{i}t)^{(n-1)(n-2)/2}(x-2\textrm{i}t)^{n(n+1)/2}+(\textrm{lower degree terms}), \\
&&\tau^{(3)}_{1}=ic_{0}^{(3)}(x+2\textrm{i}t)^{n(n-1)/2}(x-2\textrm{i}t)^{n(n+1)/2}+(\textrm{lower degree terms});
\end{eqnarray*}
but when $n$ is even, the polynomial degrees of $\tau^{(1)}_{1}$, $\tau^{(2)}_{1}$ and $\tau^{(3)}_{1}$ are lower than those of
$\tau^{(1)}_{0}$, $\tau^{(2)}_{0}$ and $\tau^{(3)}_{0}$ respectively (in both $x$ and $t$). Using these results, the boundary conditions of the three types of rogue waves given in Theorem 5 are proved.

\section{Dynamics of rogue waves}
In this section, we discuss dynamics of these rogue waves.
\subsection{Dynamics of type-I rogue waves}
The first-order type-I rogue wave is obtained by setting $n=1$ in Eq. (\ref{N-Rws1}), where the matrix elements can be obtained from Theorem 4. In this case, we find that
\begin{eqnarray*}
&& \mu^{(1)}_{1}=\sqrt{2} \left( x - 2\textrm{i}t + \textrm{i}s_{0} + \frac{1}{2}  \right),\ \nu^{(1)}_{1}= -\sqrt{2} \left( x + 2\textrm{i}t + \textrm{i}r_{0} - \frac{1}{2}  \right),\\
&& \tau^{(1)}_{0}=m^{(1)}_{1,1}=-\textrm{i}  \left[\left(x-2\textrm{i}t+\textrm{i}s_0+\frac{1}{2}\right)\left(x+2\textrm{i}t+\textrm{i}r_0+\frac{1}{2}\right)+ \right.\\
&& \hspace{2.6cm}\left. \left(x-2\textrm{i}t+\textrm{i}s_0-\frac{1}{2}\right)\left(x+2\textrm{i}t+\textrm{i}r_0-\frac{1}{2}\right) \right], \\
&& \tau^{(1)}_{1}=
\left|
\begin{array}{cc}
 m^{(1)}_{1,1} & \nu^{(1)}_{1} \\
 \mu^{(1)}_{1} & 0 \\
\end{array}
\right|= 2\left( x - 2\textrm{i}t + \textrm{i}s_{0} + \frac{1}{2}  \right)\left( x + 2\textrm{i}t + \textrm{i}r_{0} - \frac{1}{2}  \right),
\end{eqnarray*}
where $r_0$ and $s_0$ are real constants. Thus, the first-order type-I rogue-wave is
\begin{eqnarray*}
&&\hspace{1.0cm}u_1^{(1)}(x,t)=e^{-2 \mathbf{i}\hspace{0.06cm}t} \left(1+2\  \textmd{i} \frac{\tau^{(1)}_1}{\tau^{(1)}_0}\right) \\
&& \hspace{1.0cm}=e^{-2 \mathbf{i}\hspace{0.06cm}t}
\left[1-\frac{2 \left( 2x - 4\textrm{i}t + 2\textrm{i}s_{0} + 1  \right)\left( 2x + 4\textrm{i}t + 2\textrm{i}r_{0} - 1  \right)}
{16 t^2+4 x^2+8\left(r_0-s_0\right)t+4 \textmd{i} \left(r_0 +s_0\right)x -4 r_0 s_0+1}\right].
\end{eqnarray*}
Denoting $t_{0}\equiv (r_0-s_0)/4$ and $x_0\equiv (r_0+s_0)/2$, this rogue wave can be rewritten as
\begin{eqnarray}\label{rogueu11}
u_1^{(1)}(x,t)=-e^{-2\mathbf{i}\hspace{0.06cm} t}\left[1+\frac{4(4 \textrm{i} \hat{t}-1)}{16 \hat{t}^2+4 \left(x+\textrm{i} x_0\right){}^2+1}\right],\ \ \hat{t}=t+t_{0}.
\end{eqnarray}
This solution has one non-reducible real parameter $x_0$, since the parameter $t_0$ can be removed by time-translation invariance.

When $x_{0}^2 < 1/4$, this rogue wave is nonsingular. If $x_0=0$, it is the classical $x$-symmetric Peregrine solution of the local NLS equation, which is shown in Fig. 1(a) [note that any $x$-symmetric solution of the local NLS equation would satisfy the nonlocal equation (\ref{e:PTNLS})]. The peak amplitude of this Peregrine solution is 3, i.e., three times the level of the constant background. But if $x_0\ne 0$, this solution would be $x$-asymmetric and would not satisfy the local NLS equation, and its peak amplitude would become higher. One of such solutions is shown in Fig. 1(b).

\begin{figure}[htb]
\begin{center}
\includegraphics[scale=0.355, bb=0 0 385 567]{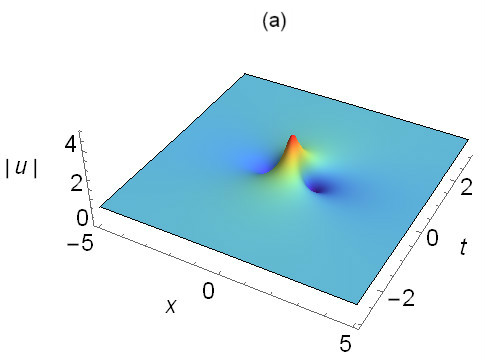}\hspace{1.2cm}
\includegraphics[scale=0.355, bb=0 0 385 567]{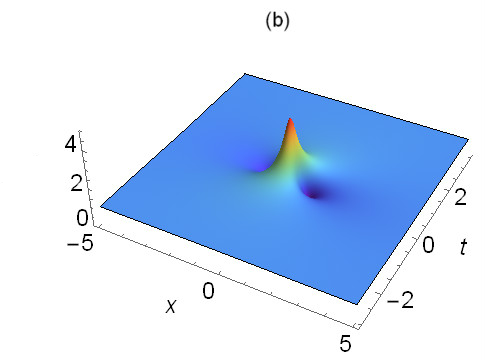}\hspace{0.2 cm}

\includegraphics[scale=0.272, bb=0 0 385 567]{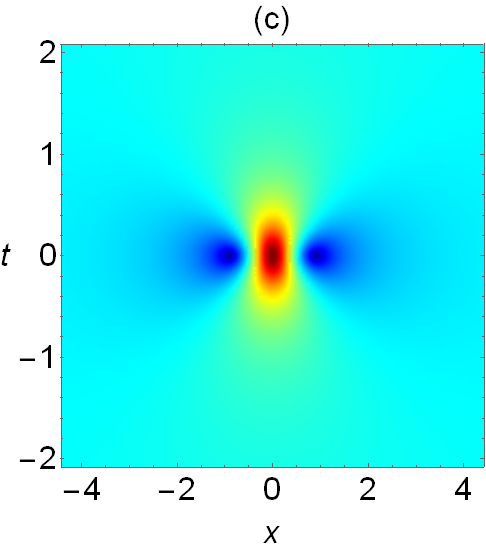}\hspace{2.2 cm}
\includegraphics[scale=0.265, bb=0 0 385 567]{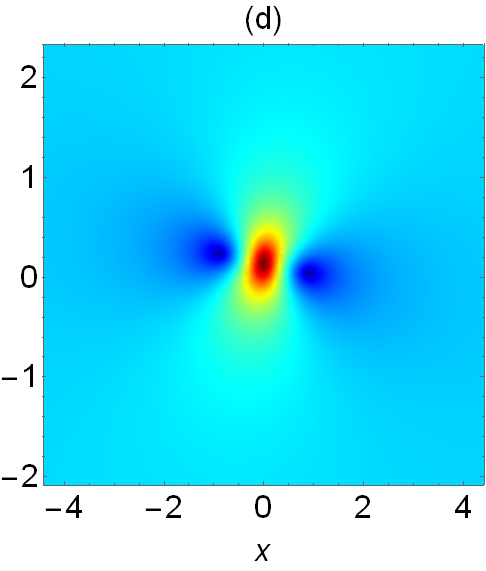}
\end{center}
\caption{Two nonsingular first-order type-I rogue waves (\ref{rogueu11}). (a) $s_0=0,\ r_0=0$ (corresponding to $x_0=0$);  (b) $s_0=1/2,\ r_0=-1/20$ (corresponding to $x_0=0.225$); (c), (d) are the corresponding density plots.}  \label{f:fig1}
\end{figure}

When $x_{0}^2\geq 1/4$, however, this rogue wave would collapse at $x=0$ and two time values $t_{c}=\pm \sqrt{(4x_{0}^2-1)/16}$.
One such solution is displayed in Fig. 2. Note that wave collapse has been reported in bright solitons of the nonlocal NLS equation (\ref{e:PTNLS}) before \cite{AblowitzMussPRL2013,AblowitzMussNonli2016}. Here we see that collapse occurs for rogue waves as well.

\begin{figure}[htb]
\includegraphics[scale=0.30, bb=0 0 385 567]{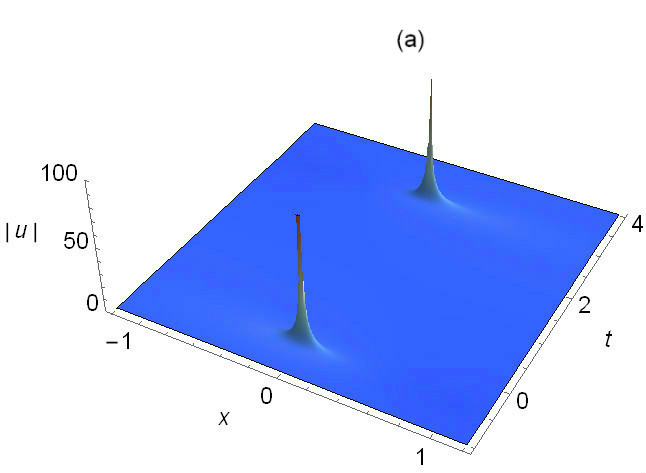}\hspace{3.2cm}
\includegraphics[scale=0.25, bb=0 0 385 567]{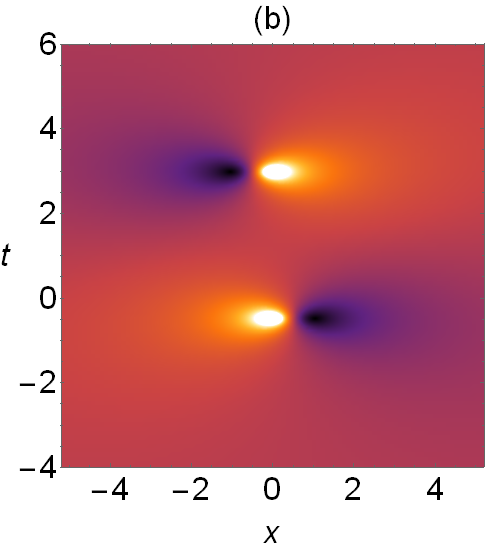}
\caption{A collapsing first-order type-I rogue wave (\ref{rogueu11}) with $s_0=6,\ r_0=1$ (corresponding to $x_0=3.5$). (a) 3D plot; (b) density plot.}   \label{f:fig2}
\end{figure}

Next, we consider second-order type-I rogue waves, which are given in Theorems 1 and 4 (with $n=2$). By tuning the free real parameters $s_0, r_0, s_1$ and $r_1$, we can get both nonsingular and singular (collapsing) solutions. Two nonsingular solutions are displayed in Fig. 3, where a single-peak pattern and a triangular pattern are observed. These patterns resemble those in the local NLS equation \cite{AAS2009,DGKM2010,DPMB2011,GLML2012,OhtaJY2012,DPMVB2013}, even though the present solutions are $x$-asymmetric and thus do not satisfy the local NLS equation.

\begin{figure}[htb]
\centering
 \includegraphics[scale=0.360, bb=0 0 385 567]{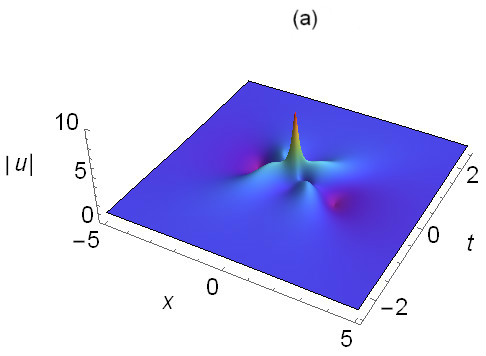}\hspace{1.3cm}
 \includegraphics[scale=0.360, bb=0 0 385 567]{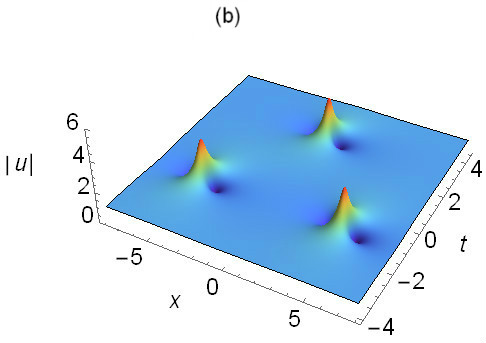}\\
 \includegraphics[scale=0.290, bb=0 0 385 567]{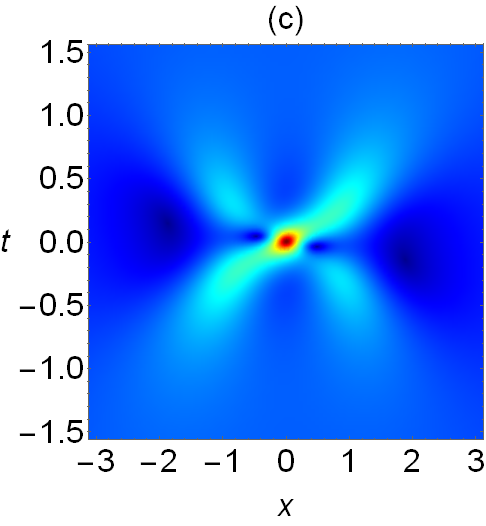}\hspace{2.0cm}
 \includegraphics[scale=0.255, bb=0 0 385 567]{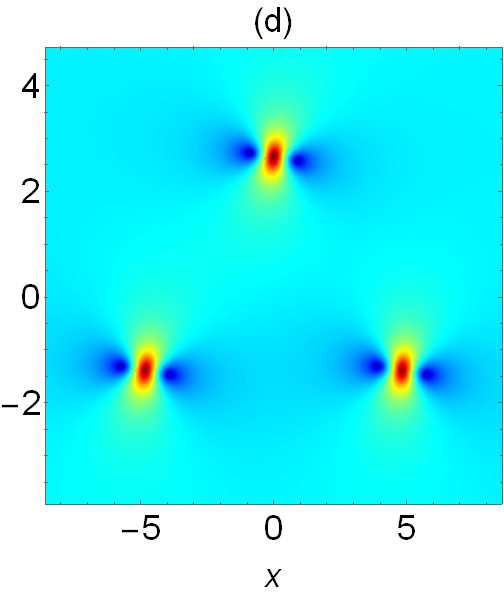}
\caption{Two nonsingular $x$-asymmetric second-order type-I rogue waves. (a) $s_0=r_0=1/6,\ s_1=r_1=0$;  (b) $s_0=r_0=1/6, s_1=100, r_1=-100$. (c), (d) are the corresponding density plots. } \label{f:fig3}
\end{figure}

More interesting are the collapsing solutions, which show more complex patterns which have not been observed before. Six of them are displayed in Fig. 4. In panel (a), the solution contains two singular (collapsing) peaks on the vertical $t$ axis, plus two ``Peregrine-like" nonsingular peaks on the horizontal $x$ axis. Panel (b) contains a quartet of singular peaks, and one ``Peregrine-like" nonsingular peak in the middle. The other four panels each contain six singular peaks, which are arranged in various circular and double-triangle patterns. Note that the maximum number of singular peaks in these solutions is six, which matches the polynomial degree of the denominator $\tau_0^{(1)}$ given in Theorem 5 at this $n=2$ value.

\begin{figure}[htb]
\begin{center}
 \includegraphics[scale=0.225, bb=0 0 385 567]{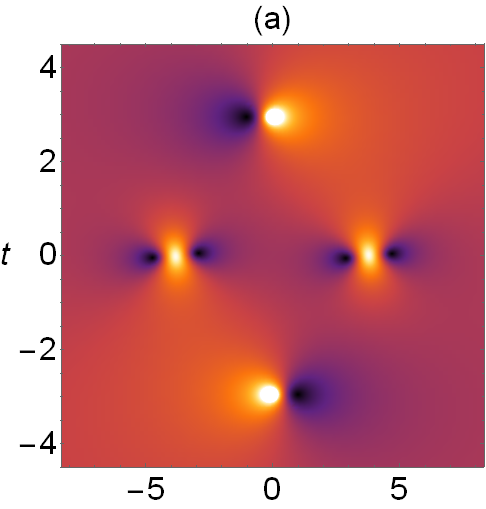}\hspace{0.85cm}
 \includegraphics[scale=0.215, bb=0 0 385 567]{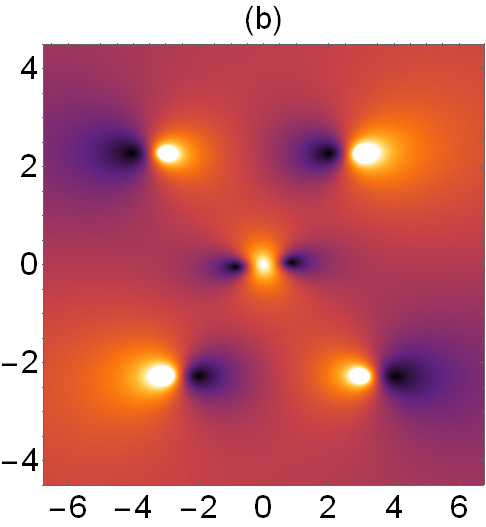}\hspace{0.85cm}
 \includegraphics[scale=0.215, bb=0 0 385 567]{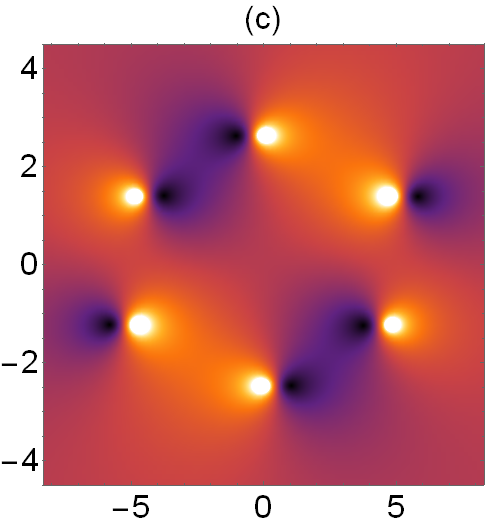}

 \includegraphics[scale=0.225, bb=0 0 385 567]{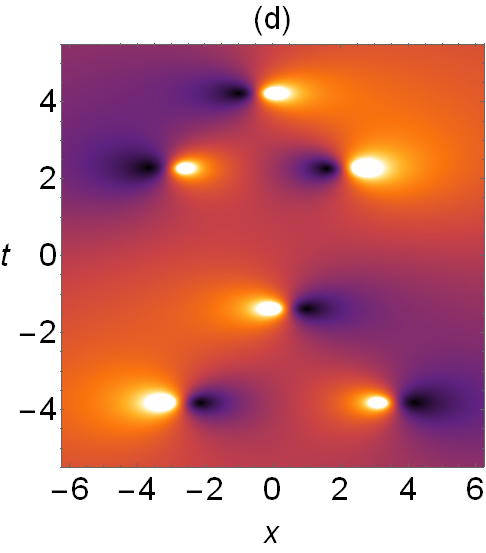}\hspace{0.85cm}
 \includegraphics[scale=0.215, bb=0 0 385 567]{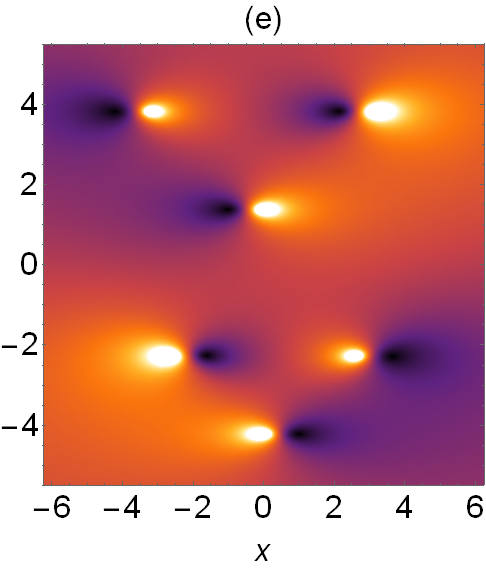}\hspace{0.85cm}
 \includegraphics[scale=0.215, bb=0 0 385 567]{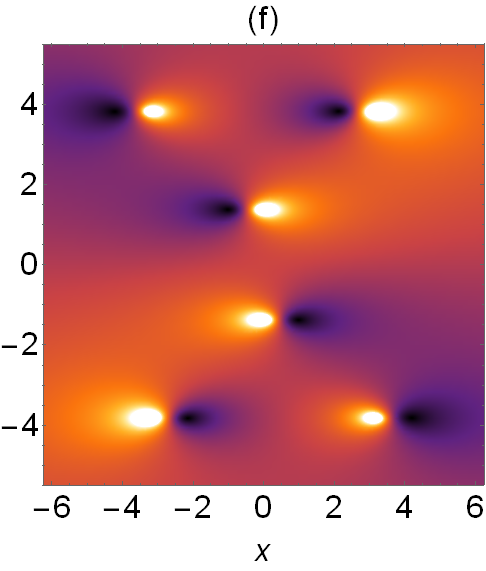}
\end{center}
\caption{Six collapsing second-order type-I rogue waves.
   (a) $s_0= r_0=2,\ s_1=r_1=50$;
   (b) $s_0= r_0=3,\ s_1=r_1=-20$;
   (c) $s_0=1/6, r_0=-1/6,\ s_1=r_1=100$;
   (d) $s_0=r_0=6,\ s_{1}=20, \ r_{1}=-20$;
   (e) $s_0=r_0=6,\ s_{1}=-20, \ r_{1}=20$;
   (f) $s_0=r_0=6,\ s_{1}=-20, \ r_{1}=-20$.} \label{f:fig4}
\end{figure}

\begin{figure}[htb]
\begin{center}
\includegraphics[scale=0.290, bb=0 0 385 567]{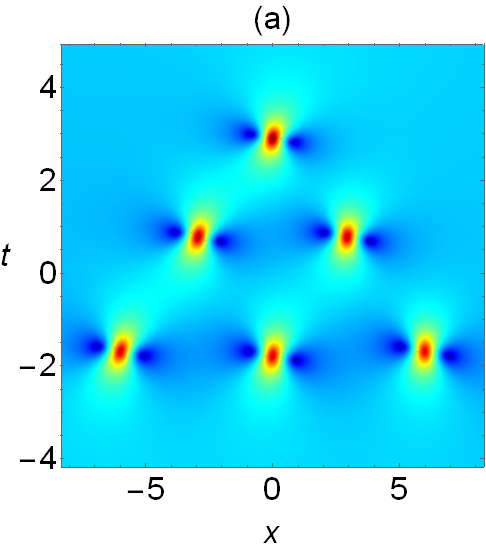}\hspace{1.4cm}
\includegraphics[scale=0.280, bb=0 0 385 567]{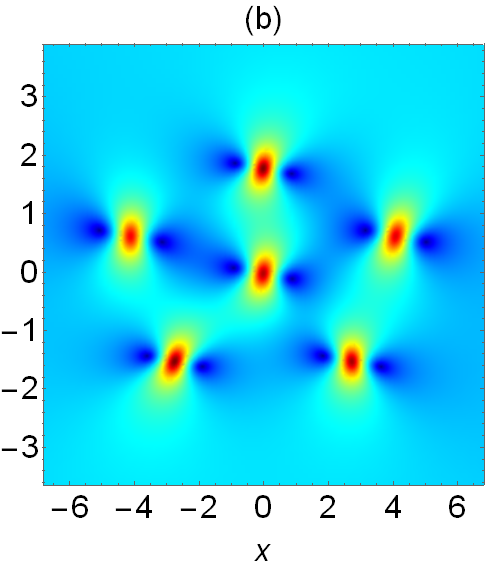}

\includegraphics[scale=0.300, bb=0 0 385 567]{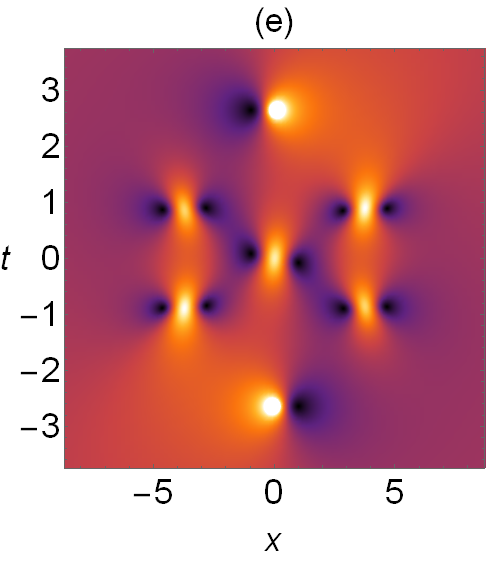}\hspace{0.90cm}
\includegraphics[scale=0.300, bb=0 0 385 567]{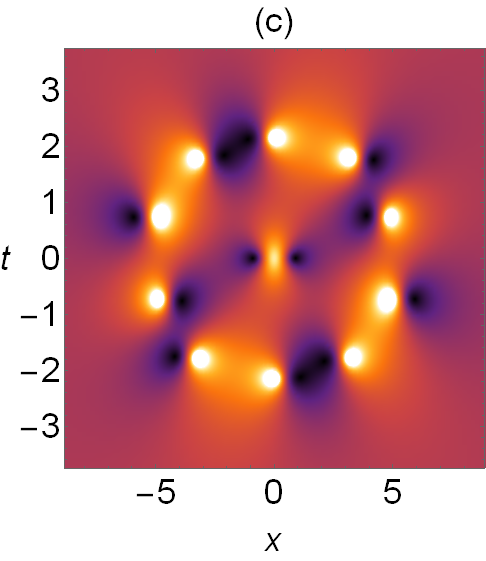}

\includegraphics[scale=0.310, bb=0 0 385 567]{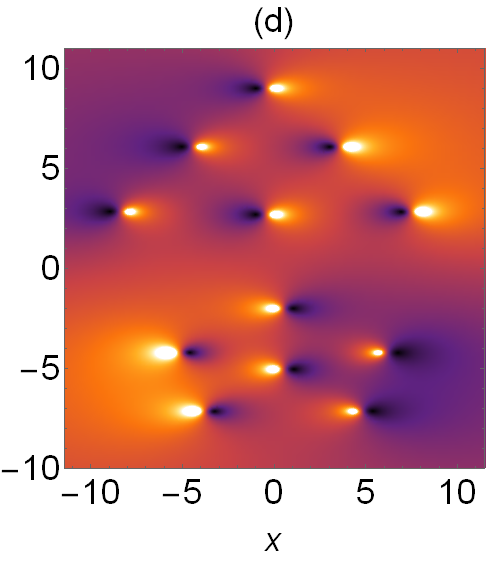}\hspace{0.90cm}
\includegraphics[scale=0.300, bb=0 0 385 567]{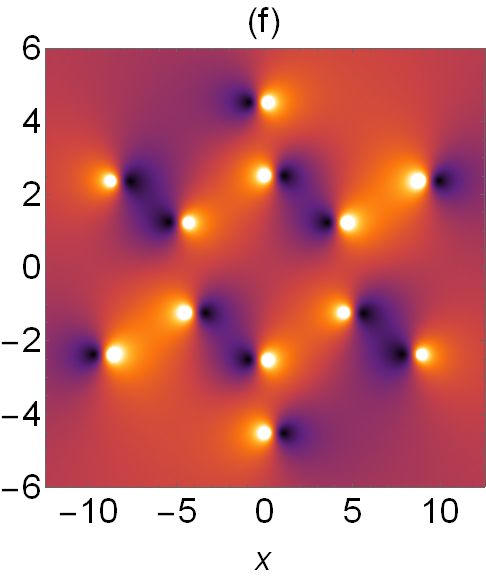}\hspace{0.90cm}
\end{center}
\caption{Six third-order type-I rogue waves (top row: nonsingular solutions; lower two rows: collapsing solutions).
(a) $s_0=1/6, r_0=1/4, s_1=30, r_1=-30, s_2=0, r_2=0$;
(b) $s_0=1/6, r_0=1/4, s_1=0, r_1=0, s_2=100, r_2=-100$;
(c) $s_0=r_0=0, s_1=0, r_1=0, s_2=r_2=240$;
(d) $s_0=r_0=10, s_1=80, r_1=0, s_2=0, r_2=-600$.
(e) $s_0=r_0=1, s_1=r_1=10, s_2=r_2=120$;
(f) $s_0=r_0=0, s_1=r_1=100, s_2=r_2=0$. } \label{f:fig5}
\end{figure}

Third-order type-I rogue waves would exhibit an even wider variety of solution patterns. Six of them are displayed in Fig. \ref{f:fig5}. The top row shows two nonsingular solutions, which contain six ``Peregrine-like" peaks arranged in triangular and pentagon patterns, reminiscent of similar solutions in the local NLS equation \cite{KAAN2011,GLML2012,OhtaJY2012,DPMVB2013}. The lower two rows show four collapsing solutions, with panel (c) containing two singular peaks and five ``Peregrine-like" nonsingular peaks in between, panel (d) containing
ten singular peaks surrounding one ``Peregrine-like" nonsingular peak, panel (e) containing twelve singular peaks in a pentagon-triangular mixed pattern, and panel (f) also containing twelve singular peaks but in a more exotic pattern. Again, this maximum number of singular peaks twelve matches the polynomial degree of the denominator $\tau_0^{(1)}$ at $n=3$.

\subsection{Dynamics of type-II rogue waves}
\begin{figure}[htb]
\begin{center}
   \includegraphics[scale=0.280, bb=0 0 385 567]{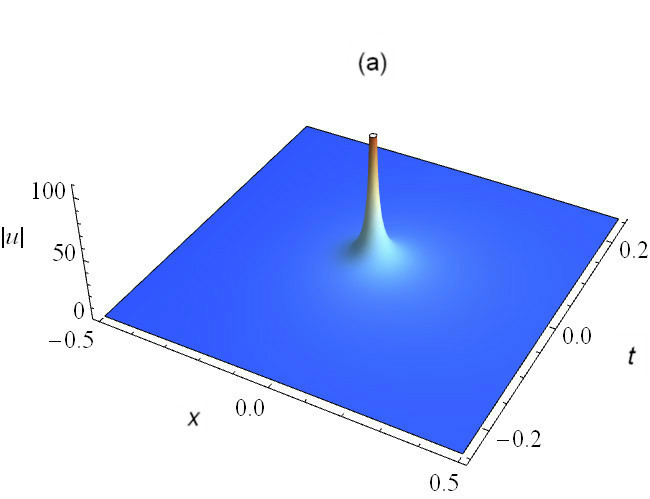}\hspace{3.5cm}
   \includegraphics[scale=0.225, bb=0 0 385 567]{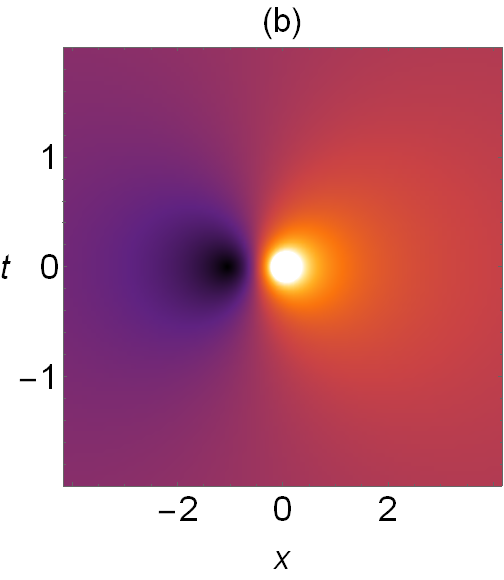}
\end{center}
\caption{The first-order type-II rogue wave (\ref{e:rogueIIn1}), with parameter $s_0=0$. (a) 3D plot; (b) intensity plot.} \label{f:fig6}
\end{figure}

Type-II rogue waves are given in Theorems 2 and 4. First, we consider the first-order of such solutions, where $n=1$. In this case, we get
\begin{eqnarray*}
&& \mu^{(2)}_{1}=\mu^{(1)}_{1},\  \nu^{(2)}_{1}=-1, \\
&& \tau^{(2)}_{0}=m^{(2)}_{1,1}=-\textmd{i}\sqrt{2}\left(x-2\textmd{i}t+\textmd{i}s_0\right),\\
&&  \tau^{(2)}_{1}=\left|
\begin{array}{cc}
m^{(2)}_{1,1} & \nu^{(2)}_{1} \\
\mu^{(2)}_{1} & 0 \\
\end{array}
\right|= \sqrt{2} \left( x - 2\textrm{i}t + \textrm{i}s_{0} + \frac{1}{2} \right);
\end{eqnarray*}
thus,
\begin{eqnarray} \label{e:rogueIIn1}
&&  u^{(2)}_1(x,t)=-e^{-2 \textmd{i} t}\left[1+\frac{1}{x-2\textmd{i}\hat{t}}\right], \quad \hat{t}=t-s_0/2.
\end{eqnarray}
Here $s_0$ is a free real parameter, which can be removed by a shift of time. This rogue wave is shown in Fig. \ref{f:fig6}. It collapses once at $x=0$ and $\hat{t}=0$. In addition, it spatially decays to the constant background in proportion to $1/x$, which is slower than the classical Peregrine solution. A counterpart of this solution in the nonlocal Davey-Stewartson equations has been reported in \cite{HePTDS,HePPTDS}.

Now we consider the second-order type-II rogue waves by setting $n=2$ in Eq.~(\ref{N-Rws2}). Using Theorem 4, we find that these solutions are given by
\begin{eqnarray}\label{Cubic-Rw1}
u_2^{(2)}(x,t)=e^{-2 \textrm{i} t}\left[1 + \frac{3 \left( 2 x-4 \textrm{i} t+2   \textrm{i} s_0+ 1\right)^2}
{4\left( x - 2\textrm{i} t + \textrm{i} s_0 \right)^3- 3\left(x-6\textrm{i} t+ \textrm{i} s_{0}+ 2 \textrm{i} s_{1}\right)}\right],
\end{eqnarray}
which can be rewritten as
\begin{eqnarray}\label{Cubic-Rw1b}
u_2^{(2)}(x,t)=e^{-2 \textrm{i} t}\left[1 + \frac{3 \left( 2 x-4 \textrm{i} \hat{t}+ 1\right)^2}
{4\left( x - 2\textrm{i} \hat{t}\right)^3- 3\left(x-6\textrm{i} \hat{t}+ 2 \textrm{i} \hat{s}_{1}\right)}\right],
\end{eqnarray}
where $\hat{t}=t-s_0/2$, and $\hat{s}_1=s_1-s_0$. This solution contains a single non-reducible real parameter $\hat{s}_1$ after the parameter $s_0$ is removed by time translation.

\begin{figure}[htb]
\begin{center}
 \includegraphics[scale=0.2575, bb=0 0 385 567]{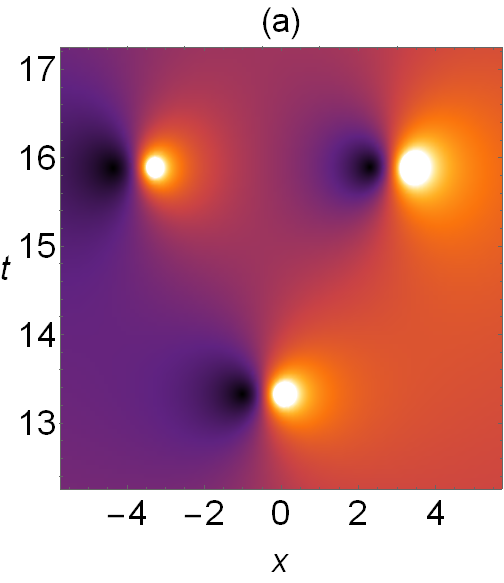}\hspace{1.65cm}
 \includegraphics[scale=0.2600, bb=0 0 385 567]{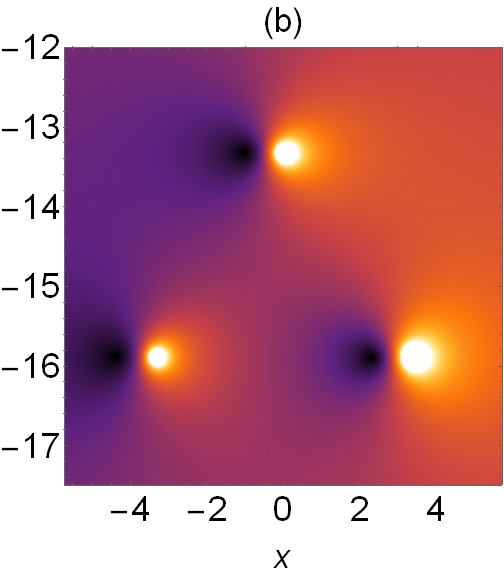}
\end{center}
\caption{Two second-order type-II rogue waves (\ref{Cubic-Rw1b}). (a) $s_0=30, s_{1}=0$; (b) $s_0=-30, s_{1}=0$.} \label{f:fig7}
\end{figure}

It is easy to show that this solution always collapses three times --- one at $x=0$, and the other two at locations symmetric in $x$. In addition, the latter two collapses occur at the same time. To illustrate, two such collapsing rogue waves are displayed in Fig. \ref{f:fig7}.

\begin{figure}[htb]
\begin{center}
 \includegraphics[scale=0.215, bb=0 0 385 567]{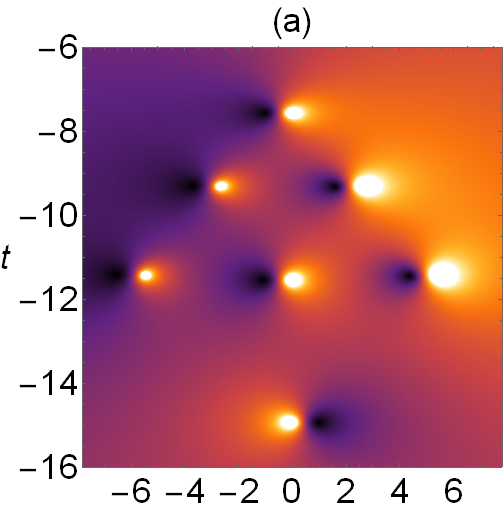}\hspace{1.1cm}
 \includegraphics[scale=0.195, bb=0 0 385 567]{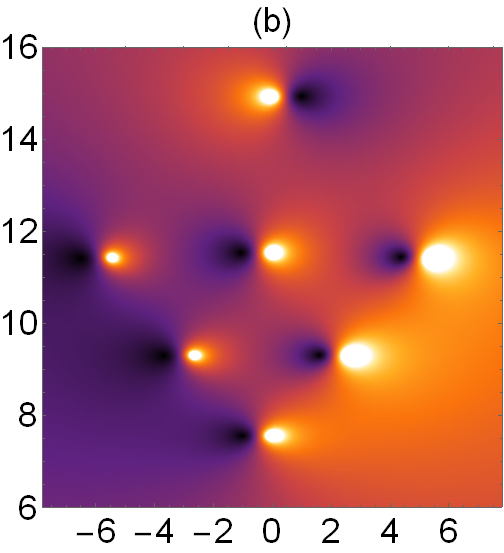}\hspace{1.1cm}
 \includegraphics[scale=0.195, bb=0 0 385 567]{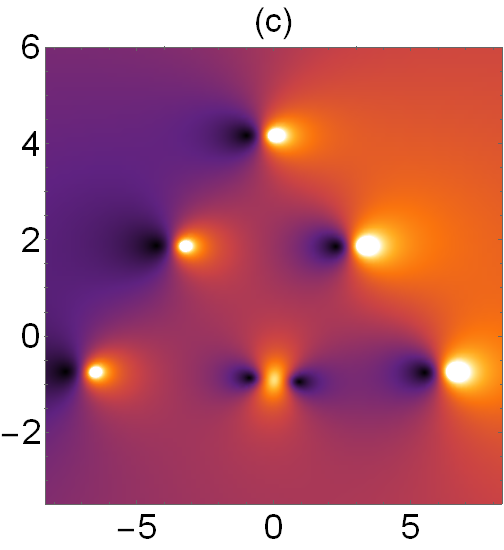}

 \vspace{0.2cm}
 \includegraphics[scale=0.208, bb=0 0 385 567]{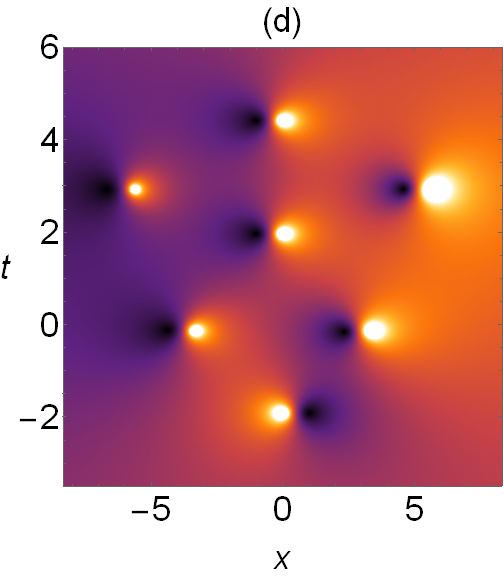}\hspace{1.1cm}
 \includegraphics[scale=0.200, bb=0 0 385 567]{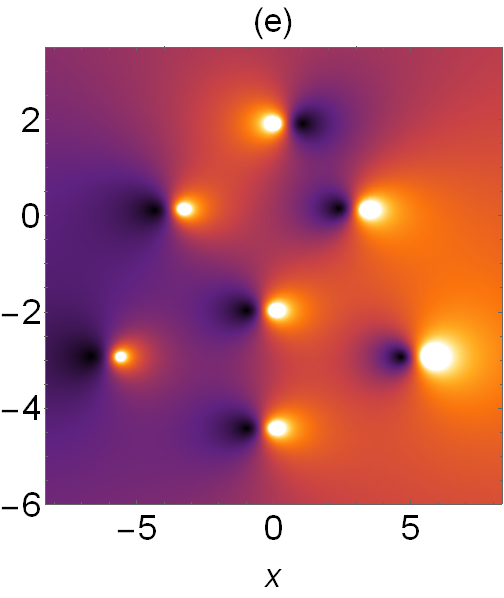}\hspace{1.1cm}
 \includegraphics[scale=0.200, bb=0 0 385 567]{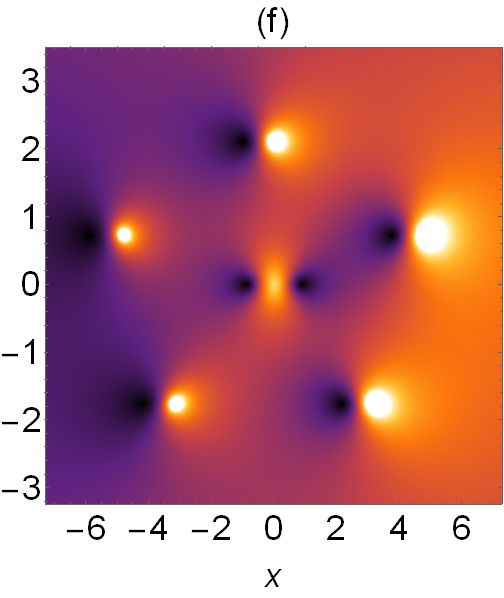}
\end{center}
\caption{Six third-order type-II rogue waves. (a) $s_0=-20,\ s_1=s_2=0,\ r_0=30$;  (b) $s_0=20, s_1=s_2=0, r_0=-30$; (c) $s_0=r_0=2$, $s_{1}=0$, $s_{2}=40$;
  (d) $s_0=r_0=4$, $s_{1}=0$,  $s_{2}=400$; (e) $s_0=r_0=-4$, $s_{1}=0$, $s_{2}=-400$; (f) $s_0=r_0=0$, $s_{1}=0$, $s_{2}=200$. } \label{f:fig8}
\end{figure}

Third-order type-II rogue waves are obtained by setting $n=3$ in Eq.~(\ref{N-Rws2}), and they contain four free real parameters, $s_0, s_1, s_2$ and $r_0$. Six of such solutions are displayed in Fig. \ref{f:fig8}. These solutions always collapse, and this collapsing exhibits various patterns such as triangles and pentagons. The maximum number of collapsing points is 7 [as in panels (a,b,d,e)], which matches the polynomial degree of $\tau_0^{(2)}$ for $n=3$. But this number of collapsing points can be less than 7 [as in panels (c,f)]; in which case pairs of collapsing points are replaced by ``Peregrine-like" nonsingular peaks.

\subsection{Dynamics of type-III rogue waves}
\begin{figure}[htb]
  \includegraphics[scale=0.230, bb=0 0 385 567]{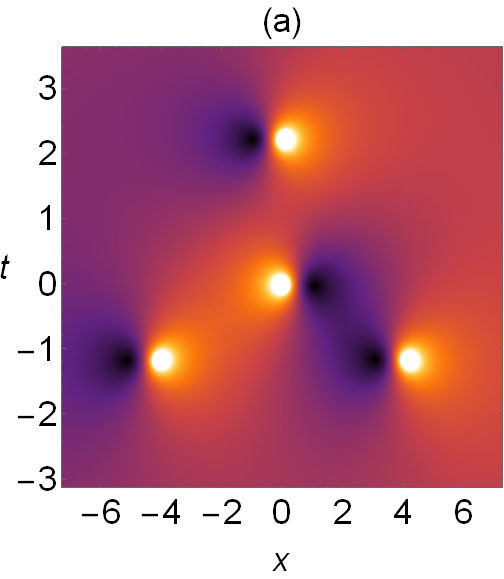}\hspace{0.85 cm}
  \includegraphics[scale=0.220, bb=0 0 385 567]{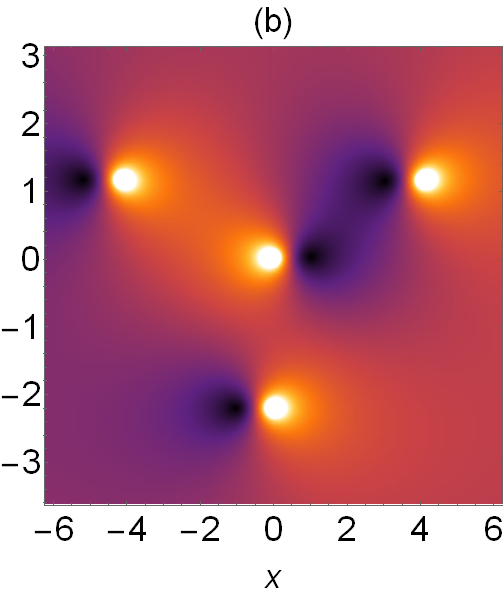}\hspace{0.85 cm}
  \includegraphics[scale=0.220, bb=0 0 385 567]{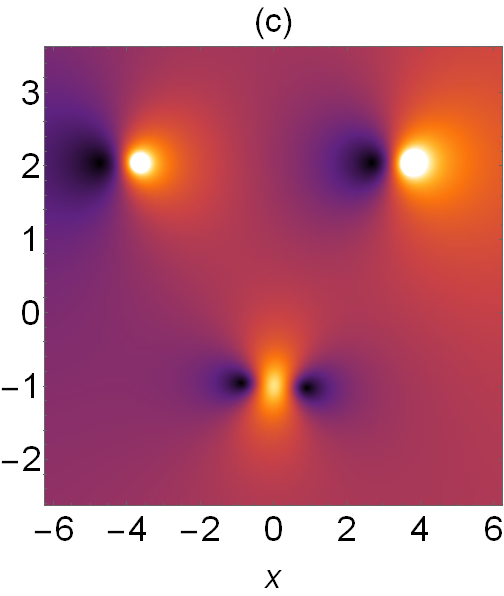}
\caption{Three second-order type-III rogue waves. (a) $s_0=0, s_{1}=60, r_{0}=r_{1}=0$; (b) $s_0=2, s_{1}=-60, r_{0}=r_{1}=0$;   (c) $s_0=2, s_{1}=-40, r_0=2, r_{1}=0$. } \label{f:fig9}
\end{figure}

Type-III rogue waves are obtained from Eq. (\ref{N-Rws3}). The first-order one, with $n=1$, turns out to be the same as the first-order type-II rogue wave given in Eq. (\ref{e:rogueIIn1}) and Fig. \ref{f:fig6}. The second-order one, with $n=2$, has the polynomial degree of its denominator $\tau_0^{(3)}$ to be 4. These rogue waves have 4 free real parameters $s_0, r_0, s_1$ and $r_1$. Three such solutions are shown in Fig. \ref{f:fig9}. The solutions in panels (a) and (b) contain four singular peaks arranged in novel patterns, while the one in panel (c) contains two singular peaks and one ``Peregrine-like" nonsingular peak.

\section{Summary and discussion}
In summary, we have derived three types of rogue waves for the focusing nonlocal NLS equation (\ref{e:PTNLS}) by Darboux transformation and Schur polynomials. The first type of $n$-th order rogue waves have denominator degrees $n(n+1)$, and can be bounded or collapsing depending on their $2n$ real parameters. The second and third types of $n$-th order rogue waves have denominator degrees $n(n-1)+1$ and $n^2$, and they appear to be collapsing for all their $2n-2$ ($n>1$) and $2n$ real parameters. These rogue waves also exhibit rich solution patterns, encompassing not only those in the local NLS equation, but also many new ones. These results reveal that the nonlocal NLS equation admits a wider variety of rogue waves, which could be useful in physical systems where this nonlocal NLS equation arises.

In the end, we should add that, by other choices of wave functions and adjoint wave functions in the Darboux transformation (see section \ref{e:section32}), we can get additional rogue wave solutions. But all additional solutions we got turn out to be equivalent to those three types reported in this article. For instance, if we choose the wave functions to be all type-b, but choose the adjoint wave functions to be all type-a, then we would get rogue wave solutions $[u_n^{(3)}(-x,-t)]^*$, which are equivalent to the type-III rogue waves in Theorem 3 since the nonlocal NLS equation (\ref{e:PTNLS}) is \PT-invariant (see Introduction). For another instance, if we choose the first wave function and adjoint wave function to be type-c, but the remaining wave functions and adjoint wave functions to be all type-a, then we would just get type-I rogue waves but with a negative sign. Whether there exists additional types of rogue waves which are not equivalent to the three types of this paper
is still an open question.

\section*{Acknowledgment}
This material is based upon work supported by the Air Force
Office of Scientific Research under award number FA9550-12-1-0244, and the National Science Foundation under award
number DMS-1616122. The work of B.Y. is supported by a visiting-student
scholarship from the Chinese Scholarship Council.

\end{document}